\documentclass[aps,prc,times,graphicx,tighten,showpacs]{revtex4}
\usepackage[dvips]{graphicx}
\usepackage[dvips]{color}
\usepackage{comment}
\newcommand{\bea}{\begin{eqnarray}}
\newcommand{\eea}{\end{eqnarray}}
\newcommand{\be}{\begin{equation}}
\newcommand{\ee}{\end{equation}}
\newcommand{\np}{{\bf p}}

\newcommand{\nr}{{\bf r}}

\newcommand{\nh}{{\bf h}}

\newcommand{\nq}{{\bf q}}

\newcommand{\Qbar}{\not{\!Q}}

\newcommand{\kbar}{\not{\!k}}
\newcommand{\Pbar}{\not{\!P}}
\newcommand{\tauvec}{\mbox{\boldmath $\tau$}}

\newcommand{\Ivec}{\mbox{\boldmath $I$}}

\newcommand{\nH}{{\bf H}}
\newcommand{\nv}{{\bf v}}
\newcommand{\nuu}{{\bf u}}
\newcommand{\grado}{\mbox{$^{\rm o}$}}

\begin{document}

\title{ 
Exploring semi-inclusive two-nucleon emission in neutrino
  scattering: a factorized approximation approach 
}

\author{
V.L. Martinez-Consentino$^a$,
J.E. Amaro$^a$,
}

\affiliation{$^a$Departamento de F\'{\i}sica At\'omica, Molecular y Nuclear,
and Instituto de F\'{\i}sica Te\'orica y Computacional Carlos I,
Universidad de Granada, Granada 18071, Spain}

\date{\today}

\begin{abstract}
 The semi-inclusive cross section of two-nucleon emission
  induced by neutrinos and antineutrinos is computed employing the
  relativistic mean field model of nuclear matter and the dynamics of
  meson exchange currents. Within this model we explore a
  factorization approximation based on the product of an integrated
  two-hole spectral function and a two-nucleon cross section averaged
  over hole pairs.  We demonstrate that the integrated spectral
  function of the uncorrelated Fermi gas can be analytically computed,
  and we derive a simple fully relativistic formula for this function,
  showcasing its dependency solely on both missing momentum and
  missing energy.  A prescription for the average momenta of the two
  holes in the factorized two-nucleon cross section is provided,
  assuming that these momenta are perpendicular to the missing
  momentum in the center-of-mass system.  The validity of the
  factorized approach is assessed by comparing it with the
  unfactorized calculation.  Our investigation includes the study of
  the semi-inclusive cross section integrated over the energy of one
  of the emitted nucleons and the cross section integrated over the
  emission angles of the two nucleons and the outgoing muon
  kinematics. A comparison is made with the pure phase-space model and
  other models from the literature. The results of this analysis offer
  valuable insights into the influence of the semi-inclusive hadronic
  tensor on the cross section, providing a deeper understanding of the
  underlying nuclear processes.
\end{abstract}


\maketitle

\section{Introduction}

The investigation of two-nucleon emission in nuclear
reactions induced by neutrinos has gained significance, particularly
in modeling the inclusive quasielastic cross section at intermediate
and high energies. Various model calculations
\cite{Mar09,Mar10,Ama11,Nie11,Cuy16,Cuy17,Roc19,Mar21b,Mar23b} have
indicated that multiparticle emission contribute
significantly, accounting for approximately 20\% or more of the
quasielastic cross section, which is primarily dominated by
one-particle emission.  Consequently, the analysis of neutrino
long-baseline experiments \cite{Gal11,Mor12,For12,Alv14,Mos16,Ath23}
requires the consideration of two-particle two-hole (2p2h)
emission events to accurately reconstruct the neutrino energy
\cite{Sob20}.

In fact, commonly used Monte Carlo event generators such as 
GENIE \cite{Dol20}, NEUT \cite{Hay09}, NUWRO \cite{Jus09,Sto17}
or GiBUU \cite{Lal12},
have incorporated the 2p2h channel from different models to
account for this contribution. Typically, these generators include
tables of the inclusive hadronic tensor \(W^{\mu\nu}(q,\omega)\) as a
function of momentum \(q\) and energy transfer \(\omega\), which are
calculated and provided by the theoretical groups. Models from
Lyon \cite{Mar09}, Valencia \cite{Nie11}, and Granada \cite{Sim17}
are currently implemented in some of these
generators, and although these models may significantly differ
depending on the kinematics,
these differences prove useful in refining the estimate of systematic
errors in Monte Carlo (MC) outputs \cite{Val06}.

The implementation of the two-nucleon emission channel requires
knowledge of the distribution of the two outgoing nucleons as
functions of their outgoing momenta \(\mathbf{p}'_1\) and
\(\mathbf{p}'_2\), for the proton-proton (pp), proton-neutron (pn),
and neutron-neutron (nn) channels. In the absence of a model for the
semi-inclusive 2p2h cross section, a first approximation is to assume
isotropic symmetry in the center-of-mass (CM) system of the outgoing
particles \cite{Sob12} when the emitted pair of nucleons absorbs
momentum \(q\) and energy \(\omega\). The corresponding distribution
is normalized using the inclusive cross section \cite{Dol20}.
However, angular symmetry in the CM is broken due to the interaction,
as the electroweak current matrix element depends non-trivially on the
moments of the initial and final particles \cite{Sim17}. To determine
the extent to which the isotropy is broken, a more realistic model for
the semi-inclusive two-nucleon emission reaction is needed, which
should be relativistic given the momenta and energies involved in
neutrino experiments, of the order of 1 GeV.

In independent-particle nuclear models, the emission of two particles
with neutrinos requires two-body current operators. These currents are
commonly modeled by assuming meson exchange between nucleons, where
the neutrino interacts with a pair of nucleons exchanging a
meson. These are known as meson exchange currents (MEC) and involve a
series of diagrams describing interactions with the exchanged meson,
possibly with the excitation of a nucleon resonance $\Delta(1232)$,
with vector and axial contributions \cite{Tow87,Ris89}.  Since the 
MEC contain the excitation of an intermediate $\Delta$, this extends
the kinematic domain of the 2p2h inclusive response, as a function of
the energy transfer, from the quasielastic peak and beyond, up to the
$\Delta$ peak \cite{Mai09}.  In more realistic nuclear models, nucleon-nucleon
correlations also allow the emission of two particles with the
one-body current, leading to interferences between the one-body and
two-body currents \cite{Roc19,Roc19b,Ben15,Ben23}.

Up to now, most models of 2p2h emission with neutrinos have focused on
calculating the inclusive reaction. The study of semi-inclusive
processes has, until recently, predominantly focused on one-particle
emission due to its major contribution to the quasielastic cross
section \cite{Fra20,Fra21,Bar21,Fra22,Fra22b,Fra23}. 
Early attempts to compute a semi-inclusive cross section with
multinucleon knockout were limited to the non-relativistic shell
model, as seen in the work of \cite{Cuy16, Cuy17}, and the calculation
presented in \cite{Sob20} using the relativistic Fermi gas with a
local density approximation. In \cite{Cuy17}, a MEC model was employed
for the two-body current, excluding the $\Delta$ excitation current,
and the final state interaction was considered with a real
single-particle potential. Meanwhile, in \cite{Sob20}, a relativistic
model based on a many-body formalism was applied, and the final-state
interaction was modeled by the cascade model implemented in the NEUT
generator.

We have recently introduced a model for semi-inclusive two-nucleon
emission induced by neutrinos \cite{Mar24}. Our approach relies on
the relativistic mean field of nuclear matter (RMF) and incorporates
relativistic MEC operators, including
seagull, pion-in-flight (pionic), pion-pole, and $\Delta$ isobar
currents. This model has been developed across a series of works
\cite{Mar21a,Mar21b,Mar23a} to compute the inclusive cross section in
the 2p2h channel, in conjunction with the superscaling approach with
relativistic effective mass (SuSAM*). Our efforts culminated in a
systematic analysis of available experimental quasielastic scattering
data of neutrinos, demonstrating reasonable agreement with the
experimental results \cite{Mar23b} similar to other different
approaches \cite{Meg13}.

 The next logical step in this framework
would be to extend the same MEC model within the RMF to predict the
semi-inclusive cross section consistently with the inclusive 2p2h
cross section. In fact, we have already applied this approach to the
semi-inclusive 2p2h reaction in \cite{Mar24} for neutrino and antineutrino scattering, where we explored the
one-fold and two-fold cross sections obtained by integrating over four or five
of the variables associated with the final momenta \(p'_1\) and
\(p'_2\).  In \cite{Mar24} we have detailed the implications of using
the RMF microscopic approach, which involves asymmetric distributions
of nucleons in the CM system of the final state, in
contrast to oversimplified modeling where isotropic distributions are
assumed. Clear differences have been observed, which should have important
implications for Monte Carlo analyses of neutrino
reactions. Additionally, focus was placed on the distributions of
proton-proton, proton-neutron  and neutron-neutron pairs, 
and, again,
important differences were observed for the microscopical 
approach versus the results found in the naive symmetric modeling.

In this work, we continue this study by analyzing other aspects of semi-inclusive cross sections for two-nucleon emission:

\begin{enumerate}
\item We will study the more general 5-folded cross section by
integrating over the energy of one of the final nucleons while keeping
constant the emission angles and the energy of the other nucleon. This
will allow us to compare with the calculations of \cite{Cuy17} in the
shell model, where MECs were considered without the $\Delta$
current. Here, we can observe the effect of MEC separately.

\item We are going to explore a factorized approximation as the
  product of a two-nucleon cross section multiplied by an integrated
  spectral function. This will allow us to see if factorized models
  developed for electron scattering from correlated nuclei, where the
  cross section is factored as the product of the one-body current by
  a combination of two-hole spectral functions, can be extended to the
  case of MEC \cite{Geu96,Ben99}. We will see that in the
  RMF, the integrated spectral function admits an analytical formula,
  simplifying the calculations. We will demonstrate that the
  two-nucleon cross section can be estimated using a prescription that
  fixes the average momenta of the holes.

\item Using the factorized formula, we will be able to calculate the
  cross section integrated over the outgoing muon and the angles of
  the final nucleons. This will allow us to compare with the
  calculation in Ref. \cite{Sob20}, where a microscopic calculation of
  this observable was performed and compared with the result from the
  NEUT event generator.

\item
 Finally, for all these observables, we will compare with the isotropic symmetric model and study the differences with our microscopic model.

\end{enumerate}

In Section 2, we summarize the formalism of semi-inclusive
two-particle emission in the RMF. In Section 3, we introduce the
factorized approximation. In Section 4, we present the results for the
5-folded cross section and for the cross section integrated over the
final muon and the nucleon angles. In Section 5, we draw our
conclusions. In the appendix, we present some mathematical details on
the derivation of the integrated two-hole spectral function.

\section{Formalism}

\subsection{Semi-inclusive 2p2h cross section}

Here we summarize the formalism used to describe the
semi-inclusive charge-changing (CC) reactions induced by neutrinos, 
\((\nu_\mu, \mu^- N_1 N_2)\), 
and antineutrinos, 
\((\overline{\nu}_\mu,\mu^+N_1N_2)\), in which two
nucleons are detected in coincidence with the muon.  The residual
daughter A-2 nucleus state is not detected. This is why we use the
convention to call this reaction {\em semi-inclusive}, in contrast to the
{\em inclusive} reaction in which only the lepton is detected, and the
{\em exclusive} reaction where the state of the daughter nucleus is also known,
and therefore, the hadronic final state is completely specified.

We closely follow the formalism of Ref. \cite{Mar24} that contains
more details on the model. The incident neutrino has four-momentum
\(k^\mu = (\epsilon, \mathbf{k})\), and the final muon has \(k'{}^\mu
= (\epsilon', \mathbf{k'})\). The energy transfer is \(\omega =
(\epsilon - \epsilon')\) and the momentum transfer is \(\mathbf{q} =
(\mathbf{k} - \mathbf{k'})\), with \(Q^2 =
\omega^2 - |\mathbf{q}|^2 < 0\). The corresponding differential 
cross-section for
detecting 
a muon with kinetic energy $T_\mu$  within a solid angle
 $\Omega_\mu=(\theta_\mu,\phi_\mu)$ and 
two nucleons with momenta \(\mathbf{p}'_1\) and
\(\mathbf{p}'_2\) can be written as
\begin{eqnarray} \label{dcros}
\frac{d\sigma_{N_1N_2}}{d T_\mu d \Omega_\mu d^3p'_1 d^3p'_2}
=
\sigma_0
L_{\mu \nu} 
W^{\mu \nu}_{N_1 N_2}(\np'_1,\np'_2,\nq,\omega)
\end{eqnarray}
where the function $\sigma_0$  is given by
\begin{equation}
\sigma_0(k,k')=
\frac{G^2 \cos^2 \theta_c}{4 \pi^2} \frac{k'}{k}
\end{equation}
In this equation the Fermi constant is $G=1.166\times 10^{-11}\quad\rm MeV^{-2}$,
and  the cosine of the Cabibbo angle is $\cos\theta_c=0.975$.

In Eq. (\ref{dcros}) the leptonic tensor, $L_{\mu \nu}$, is given by:
\begin{eqnarray}\label{leptonic}
L_{\mu \nu}&=& 
 k_\mu k'_\nu + k_\nu k'_\mu-kk'g_{\mu\nu} 
\pm i  \epsilon_{\mu \nu \alpha \beta} k^\alpha k'^\beta
\end{eqnarray}
where the sign $+(-)$ is for neutrino (antineutrino) scattering.
Finally, in Eq. (\ref{dcros}) the leptonic tensor is contracted with  
the semi-inclusive
hadronic tensor, $W^{\mu \nu}_{N_1 N_2}(\np'_1,\np'_2,\nq,\omega)$,
that contains the information about the nuclear model of
the reaction, for emitting a pair of nucleons with charges 
\(N_1,N_2\), and momenta $(\np'_1,\np'_2)$,
in an electroweak interaction that transfers energy-momentum
$(\omega,\nq)$.  In this work, we will compute
this tensor using the RMF of nuclear matter. 

In the RMF framework the nucleons
interact with scalar and vector potentials, represented as $g_s\phi_0$
and $g_vV_0$ respectively \cite{Ros80, Ser86, Weh93}.
These potentials
capture the strong interaction forces among nucleons within the
nuclear medium. The RMF model treats nucleons as interacting with
these potentials, resulting in effective masses for nucleons denoted
as $m_N^*=m_N-g_s\phi_0$. The effective mass considers the modification of the
nucleon's mass due to the scalar potential, while the vector potential
contributes a repulsive vector energy, $E_v=g_vV_0$.
In the RMF formalism the on-shell energy of a nucleon with momentum $\np$ 
is defined as
\begin{equation} \label{onshell}
E=\sqrt{p^2+m_N^{*2}}, \kern 1cm m_N^* =m_N-g_s\phi_0,
\end{equation}
while the true total  energy of the nucleon in the RMF is given by
\begin{equation}\label{energyrmf}
E_{RMF}=E+E_v=E+g_vV_0
\end{equation}
In this approach, the single nucleon states are  plane waves
$u_s(\np)e^{i\np\cdot\nr}$, where the spinor $u_s(\np)$ is a solution of
the Dirac equation with mass $m_N^*$. The ground state nuclear wave
function of the Fermi gas, $|F\rangle$, 
is constructed as a Slater determinant with all levels
occupied below some Fermi momentum $k_F$. Consequently, the action of
a two-body operator associated with the weak interaction can excite
this ground state, generating two-particle two-hole excitations (2p2h)
and leading to the emission of two particles.
\begin{equation}
|F\rangle \rightarrow |1',2',1^{-1},2^{-1}\rangle=
a^\dagger_{1'}
a^\dagger_{2'}
a_1 a_2 |F\rangle.
\end{equation}
The operators \(a_{i'}^\dagger\) and \(a_i\) are creation and
annihilation operators for single-particle states
where the states with and without prime
correspond to particles and holes, respectively, including spin
and isospin indices
\begin{equation}
|i\rangle = |\nh_i,s_i,t_i\rangle,
\kern 1cm
|i'\rangle = |\np'_i,s'_i,t'_i\rangle,
\kern 1cm
i=1,2.
\end{equation}
Applying the RMF model to the semi-inclusive two-particle emission
results in the following formula for the semi-inclusive hadronic
tensor \cite{Mar24}
 \begin{eqnarray}
W^{\mu \nu}_{N_1 N_2}(\np'_1,\np'_2,\nq,\omega)
&=& 
\frac{V}{(2\pi)^9}\int
d^3h_1
d^3h_2
\frac{(m^*_N)^4}{E_1E_2E'_1E'_2} 
 w^{\mu\nu}_{N_1N_2}(\np'_1,\np'_2,\nh_1,\nh_2)
\delta(E'_1+E'_2-E_1-E_2-\omega)
\nonumber\\
&&\times  
\delta(\np'_1+\np'_2-\nh_1-\nh_2-\nq) 
\theta(p'_1-k_F)\theta(p'_2-k_F)
\theta(k_F-h_1)\theta(k_F-h_2),
\label{hadronic}
\end{eqnarray}
where $w^{\mu\nu}_{N_1N_2}(\np'_1,\np'_2,\nh_1,\nh_2)$ represents the
elementary 2p2h hadronic tensor, and $V/(2\pi)^3 =
Z/(\frac8 3 \pi k_F^3)$ for symmetric nuclear matter. The delta
functions ensure energy-momentum conservation in the 2p2h excitation 
\begin{equation}
\np'_1+\np'_2= \nq+\nh_1+\nh_2, \kern 1cm E'_1+E'_2=\omega+E_1+E_2.
\end{equation} 
In Eq. (\ref{hadronic}) the product of step functions
impose Pauli blocking restrictions on the momenta of the particles and
holes, ensuring that the final momenta ($\np'_i$) are larger than the Fermi
momentum ($k_F$), indicating that they are excited states above the
Fermi surface, and similarly, 
the momenta of the holes ($\nh_2$) are smaller than the Fermi
momentum, indicating that they are occupied states below the Fermi
surface.

The elementary 2p2h hadronic tensor describes the transitions
between two holes and two particles
\begin{equation}
|1,2\rangle
\longrightarrow  |1',2'\rangle
\end{equation}
produced by the two-body current operator with matrix elements \cite{Ama02}
\begin{equation}
\langle 1'2' | J^\mu(\nq,\omega)| 1 2\rangle =
\frac{(2\pi)^3}{V^2}\frac{(m_N^*)^2}{\sqrt{E'_1E'_2E_1E_2}}
\delta(\np'_1+\np'_2-\nh_1-\nh_2-\nq) 
j^{\mu}(1',2',1,2),
\end{equation}
where the current functions $j^{\mu}(1',2',1,2)$ are described below.
The elementary 2p2h hadronic tensor is defined by 
\begin{eqnarray}
  w^{\mu\nu}_{N_1N_2}(\np'_1,\np'_2,\nh_1,\nh_2)
= \frac12
\sum_{s_1s_2s'_1s'_2}
j^{\mu}(1',2',1,2)^*_A \,
j^{\nu}(1',2',1,2)_A \, . 
\label{elementary}
\end{eqnarray}
where we sum over all possible spin projections of the spin-1/2
nucleons in the 2p2h excitation, as we consider the non-polarized case
where the nucleon spins are not measured. The factor $1/2$ in
Eq. (\ref{elementary}) is included to avoid double counting when
summing over spin, due to the antisymmetry of the two-body wave
function with respect to the pp or nn pair.  The two-body current matrix element is antisymmetrized
with respect to identical particles.  For the specific process
$\nu_\mu nn \rightarrow \mu^- pn$, the antisymmetrization is as follows:
\begin{equation} \label{anti}
j^{\mu}(1',2',1,2)_A=
j^\mu(1',2',1,2)- j^\mu(1',2',2,1),
\end{equation}
and for $\nu_\mu pn \rightarrow \mu^- pp$,
\begin{equation}
j^{\mu}(1',2',1,2)_A=j^{\mu}(1',2',1,2)- j^{\mu}(2',1',1,2).
\end{equation}
There are similar expressions for the antineutrino case.

\subsection{Meson-exchange currents}

\begin{figure}
\includegraphics[width=10cm, bb=120 320 500 680]{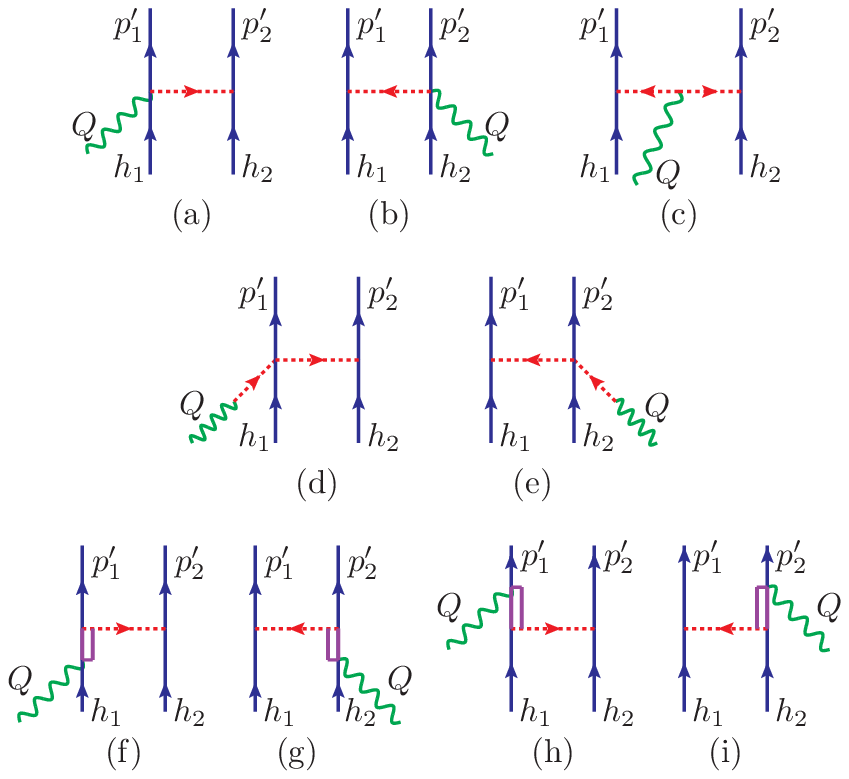}
\caption{Feynman diagrams of meson exchange currents 
considered in the present work.}
\label{diagmec}
\end{figure}

In this work, we use the electroweak MEC model described by the nine
Feynman diagrams depicted in Fig.~\ref{diagmec}. The two-body current
matrix elements \(j^{\mu}(1',2',1,2)\) corresponding to this model
enter in the calculation of the elementary 2p2h hadronic tensor,
Eq. (\ref{elementary}). The different contributions have been taken
from the pion weak production model of ref. \cite{Her07}. The MEC is
the sum of four two-body operators: seagull (diagrams a, b), pion in
flight (c), pion-pole (d, e), and \(\Delta(1232)\) excitation forward
(f, g) and backward (h, i). In the MEC model, we don't include the
correlation currents that follow from the nucleon pole diagrams of
\cite{Her07}: those currents present divergence problems due to the
double pole in the nucleon propagator \cite{Alb84,Ama10} and don't
properly account for nuclear correlations realistically because they
only involve the exchange of one pion. A more realistic description of
short-range correlations (SRC) requires using an effective
nucleon-nucleon interaction \cite{Nie11,Gra13}. Alternatively in
ref. \cite{Mar23a} a theoretical description of correlation currents
has been proposed requiring to solve the Bethe-Goldstone equation with
a realistic nucleon-nucleon interaction, a challenge that needs further work and
will be presented elsewhere. For this work, we focus on the genuine
MEC 2p2h responses to connect with our previous works on inclusive
2p2h response \cite{Mar21a,Mar21b,Mar23b}.

The relativistic MEC model for neutrino reactions was introduced in
ref. \cite{Rui17} to study the 2p2h inclusive responses in the RFG and
later extended in \cite{Mar21a,Mar21b} to the RMF, including the
effective mass and the vector energy. Following ref. \cite{Rui17}, we
explicitly separate the isospin matrix elements from the spatial and spin
dependence. This compact form will be useful, as we will see later,
for interpreting the results of semi-inclusive  pn emission.
The MEC depend on isospin across the three operators \cite{Rui17}
\begin{equation} \label{isospin}
\tauvec^{(1)}, \kern 1cm 
\tauvec^{(2)}, \kern 1cm 
\Ivec_V\equiv i \left[\tauvec^{(1)} \times\tauvec^{(2)}\right], \\
\end{equation}
that is, the isospin of the first and second particles and their vector product.
Then neutrino or antineutrino CC scattering involves the $\pm$ components
\begin{eqnarray}
\tau^{(1)}_{\pm} &=& \tau^{(1)}_{x} \pm i \tau^{(1)}_{y}  \\
\tau^{(2)}_{\pm} &=& \tau^{(2)}_{x} \pm i \tau^{(2)}_{y}  \\
I_{V\pm}  &=& (I_V)_x\pm i (I_V)_y. 
\end{eqnarray}
The MEC  can be decomposed accordingly as a sum of at most three contributions. For neutrino scattering we have 
 \begin{eqnarray}
 j^\mu_{\rm sea}
&=&
\langle t'_1t'_2| 
  I_{V+}(K_{S1}^\mu-K_{S2}^\mu) 
| t_1 t_2 \rangle,
\label{seacur}
\\
 j^\mu_{\pi}&=& 
\langle t'_1t'_2| 
I_{V+}K_{\pi}^\mu
| t_1 t_2 \rangle,
\label{picur}
\\
j^\mu_{\rm pole}
&=&
\langle t'_1t'_2| 
I_{V+} (K_{P1}^\mu-K_{P2}^\mu)
| t_1 t_2 \rangle,
\label{polecur}
\\
j^\mu_{\Delta F}
&=&
\langle t'_1t'_2| 
\frac{1}{\sqrt{6}}
\left[ 
2\tau_+^{(2)}K_{F1}
+2\tau_+^{(1)}K_{F2}
-I_{V+}(K_{F1}-K_{F2})
\right]
| t_1 t_2 \rangle,
\label{deltaF}
\\
j^\mu_{\Delta B}
&=&
\langle t'_1t'_2| 
\frac{1}{\sqrt{6}}
\left[ 
2\tau_+^{(2)}K_{B1}
+2\tau_+^{(1)}K_{B2}
+I_{V+}(K_{B1}-K_{B2})
\right]
| t_1 t_2 \rangle,
\label{deltaB}
\end{eqnarray}
and similar expressions with the minus (-) operators for antineutrino
scattering.  The nine functions $K^\mu_{S1}$, $K^\mu_{S2}$,
$K^\mu_{\pi}$, $K^\mu_{P1}$, $K^\mu_{P1}$, $K^\mu_{F1}$, $K^\mu_{F2}$,
$K^\mu_{B1}$, $K^\mu_{B2}$,
only depend on momenta and spins
$(\np'_1s'_1,\np'_2s'_2,\nh_1s_1,\nh_2s_2)$.
  They are given by
 \begin{eqnarray}
 K^\mu_{\rm S1}(\np'_1s'_1,\np'_2s'_2,\nh_1s_1,\nh_2s_2)
&=&
\frac{f_{\pi NN}   ^2}{m^2_\pi}
V_{\pi NN}^{s'_1s_1}(\np'_1,\nh_1) 
F_{\pi NN}(k_1^2)
 \bar{u}_{s^\prime_2}(\np^\prime_2)
 \left[ F^V_1(Q^2)\gamma_5 \gamma^\mu
 + \frac{F_\rho\left(k_{2}^2\right)}{g_A}\,\gamma^\mu
   \right] u_{s_2}(\nh_2),
\label{ks1}
\\
 K^\mu_{\rm S2}(\np'_1s'_1,\np'_2s'_2,\nh_1s_1,\nh_2s_2)
&=& K^\mu_{\rm S1}(\np'_2s'_2,\np'_1s'_1,\nh_2s_2,\nh_1s_1),
\label{ks2}
\\
 K^\mu_{\pi}(\np'_1s'_1,\np'_2s'_2,\nh_1s_1,\nh_2s_2)&=& 
 \frac{f_{\pi NN}^2}{m^2_\pi}
 F^V_1(Q^2)
V_{\pi NN}^{s'_1s_1}(\np'_1,\nh_1) 
V_{\pi NN}^{s'_2s_2}(\np'_2,\nh_2) 
\left(k^\mu_{1}-k^\mu_{2}\right),
\label{kpi}
\\
K^\mu_{\rm P1}(\np'_1s'_1,\np'_2s'_2,\nh_1s_1,\nh_2s_2)
&=&
\frac{f_{\pi NN}^2}{m^2_\pi}\,
\frac{F_\rho\left(k_{1}^2\right)}{g_A}
F_{\pi NN}(k_2^2)
\frac{
Q^\mu
\bar{u}_{s^\prime_1}(\np^\prime_1)\Qbar u_{s_1}(\nh_1)
 }{Q^2-m^2_\pi}
V_{\pi NN}^{s'_2s_2}(\np'_2,\nh_2), 
\label{kp1}
\\
 K^\mu_{\rm P2}(\np'_1s'_1,\np'_2s'_2,\nh_1s_1,\nh_2s_2)
&=& K^\mu_{\rm P1}(\np'_2s'_2,\np'_1s'_1,\nh_2s_2,\nh_1s_1),
\label{kp2}
\\
K^\mu_{F1}(\np'_1s'_1,\np'_2s'_2,\nh_1s_1,\nh_2s_2)
&=&
\frac{f^* f_{\pi NN}}{m^2_\pi}\,
V_{\pi NN}^{s'_2s_2}(\np'_2,\nh_2) 
F_{\pi N\Delta}(k_2^2)
\bar{u}_{s^\prime_1}(\np^\prime_1)
 k^\alpha_{2}
G_{\alpha\beta}(h_1+Q)
\Gamma^{\beta\mu}(Q)
u_{s_1}(\nh_1),
\label{kf1}
\\
 K^\mu_{\rm F2}(\np'_1s'_1,\np'_2s'_2,\nh_1s_1,\nh_2s_2)
&=& K^\mu_{\rm F1}(\np'_2s'_2,\np'_1s'_1,\nh_2s_2,\nh_1s_1),
\label{kf2}
\\
K^\mu_{B1}(\np'_1s'_1,\np'_2s'_2,\nh_1s_1,\nh_2s_2)
&=&
\frac{f^* f_{\pi NN}}{m^2_\pi}\,
V_{\pi NN}^{s'_2s_2}(\np'_2,\nh_2) 
F_{\pi N\Delta}(k_2^2)
\bar{u}_{s^\prime_1}(\np^\prime_1)
k^\beta_{2}
\hat{\Gamma}^{\mu\alpha}(Q)
G_{\alpha\beta}(p^\prime_1-Q)
u_{s_1}(\nh_1),
\label{kb1}
\\
 K^\mu_{\rm B2}(\np'_1s'_1,\np'_2s'_2,\nh_1s_1,\nh_2s_2)
&=& K^\mu_{\rm B1}(\np'_2s'_2,\np'_1s'_1,\nh_2s_2,\nh_1s_1),
\label{kb2}
\end{eqnarray}
where $k_i^\mu= (p'_i-h_i)^\mu$ is the four momentum transferred to
the $i$-th nucleon.
In these equations we have defined the following function
describing the propagation and emission (or absorption) of the exchanged pion,
\begin{equation}
V_{\pi NN}^{s'_1s_1}(\np'_1,\nh_1) \equiv 
F_{\pi NN}(k_1^2)
\frac{\bar{u}_{s^\prime_1}(\np^\prime_1)\,\gamma_5
 \kbar_{1} \, u_{s_1}(\nh_1)}{k^2_{1}-m^2_\pi},
\end{equation}
where  $F_{\pi NN}$ is a strong form factor given by
\cite{Som78,Alb84}
\begin{equation}
F_{\pi NN}(k_1^2)= \frac{\Lambda^2-m_\pi^2}{\Lambda^2-k_1^2}.
\end{equation}
with $\Lambda=1300$ MeV.
The coupling constants appearing in the currents are: $f_{\pi NN}=1$, $g_A=1.26$
and  $f^*=2.13$. 
The electroweak form factors are $F_1^V=F_1^p-F_1^n$ in the seagull
vector and pion-in flight currents, for which we use Galster parametrization,
and the $\rho$ form factor $F_{\rho}$ in the axial seagull and
pion-pole currents, is taken from \cite{Her07}.  

In the case of the forward $\Delta$ current $\Gamma^{\beta\mu}$ is
the $N\rightarrow\Delta$ transition vertex 
\begin{equation} \label{gamma-delta}
\Gamma^{\beta\mu}(Q)=
\frac{C^V_3}{m_N}
\left(g^{\beta\mu}\Qbar-Q^\beta\gamma^\mu\right)\gamma_5
+ C^A_5 g^{\beta\mu},
\end{equation}
and for the backward current
\begin{equation}
\hat{\Gamma}^{\mu\alpha}(Q)=\gamma^0
\left[\Gamma^{\alpha\mu}(-Q)\right]^{\dagger}
\gamma^0 \,.
\end{equation}
We use the vector and axial form factors in the $\Delta$ vertices from ref.
\cite{Her07}
\begin{equation}
C_3^V(Q^2)= \frac{2.13}{(1-Q^2/M_V^2)^2}\frac{1}{1-\frac{Q^2}{4 M_V^2}},
\kern 1cm
C_5^A(Q^2)= \frac{1.2}{(1-Q^2/M_{A\Delta}^2)^2}\frac{1}{1-\frac{Q^2}{4 M_{A\Delta}^2}},
\end{equation}
with $M_V=0.84$ GeV, and $M_{A\Delta}=1.05$ GeV.
In the $\Delta$ current, strong form factors are also applied.
We use the $\pi N \Delta$ strong form factor from Ref. \cite{Dek94}
\begin{equation}
F_{\pi N\Delta}(k_2^2)=
 \frac{\Lambda^2_\Delta}{\Lambda_\Delta^2-k_2^2}
\end{equation}
were $\Lambda_\Delta=1150$ MeV. 

Finally
the $\Delta$-propagator,
including the $\Delta$ decay width
is given by
\begin{equation}\label{delta_prop}
 G_{\alpha\beta}(P)= \frac{{\cal P}_{\alpha\beta}(P)}{P^2-
 M^2_\Delta+i M_\Delta \Gamma_\Delta(P^2)+
 \frac{\Gamma_{\Delta}(P^2)^2}{4}} \, ,
\end{equation}
and the projector  ${\cal P}_{\alpha\beta}(P)$ over 
spin-$\frac32$ on-shell particles is given by
\begin{eqnarray}
{\cal P}_{\alpha\beta}(P)&=&-(\Pbar+M_\Delta)
\left[g_{\alpha\beta}-\frac13\gamma_\alpha\gamma_\beta-
\frac23\frac{P_\alpha P_\beta}{M^2_\Delta}\right.
+\left.
\frac13\frac{P_\alpha\gamma_\beta-
P_\beta\gamma_\alpha}{M_\Delta}\right].
\end{eqnarray}

In the RMF the spinors \(u_{s'_i}(\np'_i)\) and \(u_{s_i}(\nh_i)\) are
the solutions of the Dirac equation with relativistic effective mass
\(m_N^*\), and with on-shell energy, Eq. (\ref{onshell}). However the
total nucleon energy in the RMF (\ref{energyrmf}) includes the vector
energy, $E_v$.  This vector energy cancels out in the terms of the
currents that depend on the vectors $k_i^\mu$. But it is not canceled
in the $\Delta$-propagator, which is the only place where $E_v$ appears
explicitly \cite{Mar21a,Mar21b}.

In this work, we do not include medium corrections to the intermediate
\(\Delta\) particle. In refs. \cite{Mar21b,Mar23b}, we studied the
effect of considering the interaction of the $\Delta$ with the RMF using
an effective mass and a vector energy for the $\Delta$. It was found that
the effect of this interaction significantly modifies the inclusive
response. However, the properties of the $\Delta$ in the medium have
uncertainties and are not unambiguously determined. Therefore, in the
absence of a definitive theory, these studies serve as a measure of
the uncertainty in the MEC response, among the many that exist. Thus,
in this work, the calculations will be done with the properties of the
$\Delta$ in vacuum, which is consistent with the inclusive responses of
refs. \cite{Mar21a,Mar21b,Mar23b}.

\subsection{Semi-inclusive hadronic tensor}

The calculation of the semi-inclusive hadronic
tensor of Eq. (\ref{hadronic}) first requires evaluating the elementary
2p2h hadronic tensor, Eq. (\ref{elementary}), by performing the sums
over spin. As in previous works \cite{Sim17}, these sums are computed
numerically because the analytical calculation in terms of traces of
gamma matrices is extremely cumbersome and not practical, and it does
not provide any advantage in terms of computation time. On the other
hand, the integration over holes in Eq. (\ref{hadronic}) can be
reduced to a two-dimensional integral due to the Dirac delta functions
of energy and momentum. In our case, the integration of the energy
delta is performed in the center-of-mass system of the two
initial particles, where the problem is reduced to an integral over
the relative angles of the hole pair.

Considering a semi-inclusive event where $\np'_1,\np'_2,\nq,\omega$ are known,
 we can compute the total momentum and energy of the two holes
 \begin{eqnarray} 
 \nH &=& \np_1'+\np_2'-\nq,\\
  E & =&  E_1'+E_2'-\omega. 
\end{eqnarray}
Then the semi-inclusive 2p2h hadronic tensor can be written
 \begin{eqnarray}
W^{\mu \nu}_{N_1 N_2}(\np'_1,\np'_2,\nq,\omega)
&=& 
\theta(p'_1-k_F)\theta(p'_2-k_F)
\frac{V}{(2\pi)^9}
\frac{(m^*_N)^4}{E'_1E'_2} 
\int
\frac{d^3h_1}{E_1}
\frac{d^3h_2}{E_2}
 w^{\mu\nu}_{N_1N_2}(\np'_1,\np'_2,\nh_1,\nh_2)
\nonumber\\
&&\times  
\delta(E_1+E_2-E)
\delta(\nh_1+\nh_2-\nH) 
\theta(k_F-h_1)\theta(k_F-h_2),
\label{hadronic2}
\end{eqnarray}
The deltas of energy and momentum inside the integral imply that only
the holes such that $\nh_1+\nh_2=\nH$ and $E_1+E_2=E$ contribute to the
integral. Therefore, we can perform the integral by going to the
center of mass of the two holes that moves with velocity $\nv=\nH/E$. 
We denote with double prime the
coordinates in the CM. Then, in this system, $\nH''=\nh''_1+\nh''_2=0$
and $E''=\sqrt{E^2-H^2}$. The two holes move back to back,
$\nh''_2=-\nh''_1$, and have the same energy $E''_1=E''_2=E''/2$. In
Appendix \ref{appa}, we demonstrate in detail how the integral transforms when
moving to the center of mass through a boost (change of
variables). Applying these results to the case of the hadronic tensor,
we can write:
 \begin{eqnarray}
W^{\mu \nu}_{N_1 N_2}(\np'_1,\np'_2,\nq,\omega)
&=& 
\theta(E^2-H^2-4m_N^{*2})\theta(p'_1-k_F)\theta(p'_2-k_F)
\frac{V}{(2\pi)^9}
\frac{(m_N^*)^4 }{E'_1 E'_2}
\nonumber\\
&&\mbox{}\times  
\frac{ h_1''}{2E_1''}
\int
d \Omega_1''
 w^{\mu\nu}_{N_1N_2}(\np'_1,\np'_2,\nh_1,\nh_2)
\theta(k_F-h_1)\theta(k_F-h_2),
\label{hadronic3}
\end{eqnarray}
where $d\Omega_1''=d\cos\theta''_1d\phi''_1$ and $\theta''_1,\phi''_1$
are the angles of the first hole in the CM system.  Note that the
integral is performed over relative angles in the CM system of the two
holes, but the momenta $\nh_1, \nh_2$ in the elementary 2p2h hadronic
tensor and in the step functions are evaluated in the Lab system. 
The steps to
calculate the integral are the following:
\begin{enumerate}
\item  First, calculate $(\nH, E)$ from $\np'_1,
\np'_2, \nq, \omega$. 
\item 
Then, calculate the holes energy in the CM,
$E''_1=E''/2=\sqrt{E^2-H^2}/2$. 
\item Next, for each value of the angles, construct the
vector 
\begin{equation}
\nh''_1=h''_1
(\cos\phi''_1\sin\theta''_1,\sin\phi''_1\sin\theta''_1,\cos\theta''_1). 
\end{equation}
\item 
 Apply an inverse boost to the laboratory system
to calculate $\nh_1$. 
\item  Calculate $\nh_2=\nH-\nh_1$. 
\end{enumerate}

The boost is performed as follows.
The CM frame is characterized by a velocity $\nv=\nH/E$, where the unit
vector $\nuu=\nv/v$ specifies its direction. To transform a CM vector
$(E''_1,\nh''_1)$ to the Lab system, we employ the Lorentz factor
$\gamma=1/\sqrt{1-v^2}$ and perform the following boost:
\begin{eqnarray}
h_{1u} &=& \gamma(vE''_1+h''_{1u})  \label{boost} \\
\nh_{1\perp} &=& \nh''_{1\perp}
\end{eqnarray}
Here, $h_{1u}= \nh_1\cdot\nuu$ is the component of $\nh_1$ along the
direction of $\nuu$, while $\nh_{1\perp}$ denotes the component
perpendicular to $\nuu$, which is an invariant quantity under the boost. 
Therefore  $\nh_1= h_{1u}\nuu + \nh_{1\perp}$, and we can compute it as follows:
\begin{eqnarray}
\nh_1 &=& \gamma(vE''_1+h''_{1u})\nuu +\nh''_{1\perp} \nonumber \\
&=& (\gamma vE''_1+(\gamma-1)h''_{1u})\nuu +\nh''_1.
\end{eqnarray}

The derived equation (\ref{hadronic3}) represents the key outcome for
the semi-inclusive 2p2h hadronic tensor, expressed as a
two-dimensional integral over relative angles, necessitating numerical
methods for evaluation. This concise formula encapsulates the exact
hadronic tensor within the RMF or the RFG when the mean field is
disconnected. In the results section, we showcase outcomes and conduct
comparisons with the factorized approximation introduced in the
subsequent section, shedding light on the intricate dynamics of
semi-inclusive two-nucleon emission reactions in neutrino scattering.

\section{Factorization of the semi-inclusive 2p2h hadronic tensor}

In this section we introduce a factorized approximation for the
semi-inclusive two-nucleon emission response. While we have developed
an exact model for the semi-inclusive hadronic tensor, represented by
a straightforward two-dimensional integral of the elementary 2p2h
hadronic tensor, a factorized approximation can prove beneficial under
certain circumstances. For instance, when calculating observables
integrated over the angles of the outgoing nucleons and the outgoing
muon, an eight-dimensional integral would be required, demanding more
intensive computational efforts. This is particularly significant as
the elementary 2p2h tensor needs to be evaluated within the integral
for all contributing events. Therefore, in this work, we aim to
investigate the validity of a factorized approximation. In this
approach, the elementary tensor
\(w^{\mu\nu}(\np'_1,\np'_2,\nh_1,\nh_2)\) is factorized by evaluating
it at averaged values for the two holes. Additionally, this
exploration connects with other formalisms describing two-nucleon
emission. For example, in reactions like $(e,e'pp)$ in presence of
correlations, in a plane wave approximation, the exclusive cross
section of factorizes as the product of a single-nucleon cross section
multiplied by a combination of two-hole (2h) spectral functions
\cite{Geu96,Ben99}.  We aim to determine if a similar factorized
approximation can be applied in the case of MEC. However, as MEC is produced by a two-body current, it is not
clear whether a unique two-nucleon cross section that factorizes
unequivocally exists. Thus, our investigation serves as a preliminary
exploration of this intriguing possibility.

\subsection{The integrated two-hole spectral function}

First, we will show that an exact factorized formula can be obtained
by defining an average of the elementary 2p2h hadronic tensor from
Equation (\ref{hadronic2}). Indeed, we first define the function
\begin{equation}
G(E,H) = 
\int d^3 h_1 d^3 h_2 \frac{(m_N^*)^2}{E_1 E_2}  
\theta(k_F-h_1) \theta(k_F-h_2)
\delta(E_1+E_2-E)
\delta(\nh_1+\nh_2-\nH).
\label{g1}
\end{equation}
Now we can use this function to define an averaged value of the elementary 2p2h 
hadronic tensor as follows
\begin{equation}
\langle w^{\mu\nu}_{N_1N_2}(\np'_1,\np'_2,\nq,\omega)\rangle\equiv
\frac{1}{G(E,H)} \int d^3 h_1 d^3 h_2 \frac{(m_N^*)^2}{E_1 E_2}  
w^{\mu\nu}_{N_1N_2}(\np'_1,\np'_2,\nh_1,\nh_2)
\theta(k_F-h_1) \theta(k_F-h_2)
\delta(E_1+E_2-E)
\delta(\nh_1+\nh_2-\nH) ,
\label{waveraged}
\end{equation}
where, as before, $E=E'_1+E'_2-\omega$, and $\nH=\np'_1+\np'_2-\nq$.
With this definition we can write the semi-inclusive two-nucleon
hadronic tensor, Eq. (\ref{hadronic2}), in  exact factorized form
 \begin{equation} \label{exact}
W^{\mu \nu}_{N_1 N_2}(\np'_1,\np'_2,\nq,\omega)
= 
\theta(p'_1-k_F)\theta(p'_2-k_F)
\frac{V}{(2\pi)^9}
\frac{(m^*_N)^2}{E'_1E'_2} 
\langle w^{\mu\nu}_{N_1N_2}(\np'_1,\np'_2,\nq,\omega)\rangle
G(E,H)
\end{equation} 
The function \(G(E,H)\) holds significant physical meaning as it is
intricately connected to the 2h spectral function within
the Fermi gas model. In the non-relativistic context, this spectral
function is given by \cite{Ben99} 
\begin{equation}
S_{2hFG}(\nh_1,\nh_2,E_m)=
\theta(k_F-h_1)
\theta(k_F-h_2)
\delta(E_m+\frac{h_1^2}{2m_N}+\frac{h_2^2}{2m_N})
\end{equation}
 where \(E_m\) represents the missing energy $E_m=\omega - T'_1 -
 T'_2$ and $T'_i$ are the kinetic energy of the final particles.  Is
 is clear that in the non relativistic case the total kinetic energy
 of the holes is minus the missing energy $E=T_1+T_2=-E_m$.  The
 association of the \(G(E,H)\) function with the 2h spectral
 function becomes evident, establishing a clear link between the
 two. This connection is rooted in the integral of the 2h spectral
 function, Eq. (\ref{g1}) subject to the constraint \(\nh_1+\nh_2=\nH\). 

\subsection{Factorized approximation}

The exact factorization expressed in Eq. (\ref{exact}) is not
practically applicable, as the calculation of the averaged elementary
2p2h tensor still requires an exact computation. The factorized
approximation assumes that we can approximate this average by
evaluating the tensor at specific hole momenta, $\langle\nh_1\rangle$ and 
$\langle\nh_2\rangle$,
representing average values for the holes involved in the
reaction. 
\begin{equation}
\langle w^{\mu\nu}_{N_1N_2}(\np'_1,\np'_2,\nq,\omega)\rangle
\simeq
w^{\mu\nu}_{N_1N_2}(\np'_1,\np'_2,\langle\nh_1\rangle,\langle\nh_2\rangle)
\end{equation}
Then the factorized formula for the semi-inclusive two-nucleon hadronic tensor reads
 \begin{equation}\label{factorized}
W^{\mu \nu}_{N_1 N_2}(\np'_1,\np'_2,\nq,\omega)
\simeq 
\theta(p'_1-k_F)\theta(p'_2-k_F)
\frac{V}{(2\pi)^9}
\frac{(m^*_N)^2}{E'_1E'_2} 
w^{\mu\nu}_{N_1N_2}(\np'_1,\np'_2,\langle\nh_1\rangle,\langle\nh_2\rangle)
G(E,H)
\end{equation}
This introduces a simplification that, if valid, could
streamline the calculation while providing valuable insights into the
semi-inclusive two-nucleon emission process induced by neutrinos.

For the factorized approximation to be useful, it is crucial to find a
prescription for the averaged hole momenta that is suitable. A
mandatory requirement is that these moments must comply with
energy-momentum conservation. This implies that the frozen
approximation, assuming $h_1=h_2=0$, cannot be taken, as these values
may not hold for all kinematics. Therefore, we turn to the results of
the previous section, where it is described how the vector $\nh_1$ is
constructed through a boost from the CM system. Indeed, we have seen
that, given $E$ and $\nH$, the value of $h''_1$ in the CM system is
fixed, as its energy is $E''_1=E''/2= \sqrt{E^2-H^2}$. The only
remaining specification is the angles $\theta''_1,\phi''_1$ of
$\langle\nh''_1\rangle$ in the CM system, followed by the boost back to the Lab
system. This procedure ensures that the averaged hole momenta are
consistent with energy-momentum conservation, providing a viable
approach to implement the factorized approximation.

To define the angles of $\langle\nh''_1\rangle$, it is necessary to establish a
reasonable prescription or algorithm, followed by a posteriori
validation through comparison with the exact result. A sensible
prescription is to choose the vector $\langle\nh''_1\rangle$ in the CM system so
that it is perpendicular to both $\nH$ and $\nq$. This approach is
based on geometric and symmetry considerations explained next.

\subsection{Prescription for $\langle\nh_1\rangle$}

In Eq. (\ref{boost}), the value of \(h_{1u}\) is obtained through the
boost from the CM to the Lab. Let's write the corresponding equation
for the energy \(E_1\) provided by the Lorentz transformation.
\begin{equation}
E_1 = \gamma(E''_1+vh''_{1u})  \label{boost2}. 
\end{equation}
On the other hand, the energy of the second hole can be obtained
by replacing \(h''_{1u}\) with \(h''_{2u} = -h''_{1u}\) because in the
CM, the two holes are moving back-to-back.
\begin{equation}
E_2 = \gamma(E''_1-vh''_{1u})  \label{boost3}. 
\end{equation}
Both energies \(E_1\) and \(E_2\) must be below the Fermi energy
\(E_F\) for \(\mathbf{h}_1\) and \(\mathbf{h}_2\) to contribute to the
semi-inclusive response. In other words, these two inequalities must be
satisfied simultaneously.
\begin{eqnarray}
 \gamma(E''_1+vh''_{1u}) \leq E_F, \\
 \gamma(E''_1-vh''_{1u}) \leq E_F, 
\end{eqnarray}
From where 
\begin{equation}
|h''_{1u}| \leq \frac{E_F-\gamma E''_1}{ \gamma v}.
\end{equation}
This inequality establishes the possible values of the component
\(h''_{1u}\) in the direction of \(\nuu=\nH/H \) in the CM system that
contribute to the semi-inclusive hadronic tensor. Alternatively, if
\(\theta\) is the angle between \(\nuu\) and \(\mathbf{h}''_1\), we have
a maximum possible value for \(|\cos\theta|\).
\begin{equation}
|\cos\theta| \leq \frac{E_F-\gamma E''_1}{ \gamma vh''_1}.
\end{equation}
 Then it follows that the average value of \(\cos\theta\) is
 \(\langle\cos\theta\rangle=0\), and this gives \(\langle h''_{1u} \rangle=0\). That is,
 \(\mathbf{h}''_1\) is perpendicular to \(\nuu\), i.e., perpendicular
 to \(\mathbf{H}\). 
Besides, the above inequality provides a condition
 for the existence of solutions, which is that the
 right-hand side must be positive, 
\begin{equation} \label{condicion}
0 \leq E_F-\gamma E''_1,
\end{equation} 
otherwise, the semi-inclusive
 hadronic tensor is zero. But
\begin{equation}
\gamma E''_1 = \frac12 \gamma \sqrt{E^2-H^2} =  
\frac12 \gamma E \sqrt{1-H^2/E^2} =  
\frac12 \gamma E \sqrt{1-v^2} = \frac{E}{2}
\end{equation}
Therefore the condition (\ref{condicion}) is equivalent to 
\begin{equation}
E < 2E_F
\end{equation} 
Therefore, the condition for there to be allowed values of \(\nh_1\) is
that the sum of the energy of the two holes, given
by \(E'_1 + E'_2 - \omega\), must be less than twice the Fermi energy. This
is consistent with the model, and if it holds, solutions
perpendicular to \(\mathbf{H}\) in the CM are always possible. This
gives consistency to the factorized approximation, as the prescription
for \(\langle\mathbf{h}'_1\rangle\) 
will always provide solutions compatible with
energy-momentum conservation and below the Fermi level. On the other
hand, in the case where \(E > 2E_F\), both the integrated spectral
function and the hadronic tensor are zero in this model, and
therefore, it is not necessary to impose the condition explicitly in the
factorized formula, as it is implicitly included in the function
\(G(E,H)\).

Still, we need to specify a choice for the perpendicular component
\(\nh''_{1\perp}\) since all possibilities are valid and compatible with
energy and momentum conservation. We must then turn to symmetry
considerations, leading us to choose \(\nh''_{1\perp}\) in such a way
that it is perpendicular to both \(\mathbf{H}\) and \(\mathbf{q}\)
simultaneously, i.e., in the direction of \(\nuu \times
\mathbf{q}\). Then our prescription is
\begin{eqnarray}
\langle \nh''_1 \rangle = h''_1 \frac{\nH\times\nq}{Hq}
\end{eqnarray}
The reason for this choice lies in the fact that inclusive responses
are known to have azimuthal symmetry around \(\mathbf{q}\) when it is
chosen along the z-axis.  Here, geometrical intuition guides us based
on the search for a similar symmetry with respect to \( \mathbf{q} \)
in the semi-inclusive hadronic tensor, so that \(
\langle\nh''_1\rangle \) and \( \langle\nh''_2\rangle \), which would
be symmetrically placed with respect to \( \mathbf{q} \), are both
perpendicular to \( \mathbf{q} \).

\subsection{Analytical form of $G(E,H)$}

The interpretation of the function \(G\) as an integrated 2h spectral
function in the RFG, and extended to the RMF, deepens our insight into
the characteristics of the emitted nucleon pair. In the factorized
approximation, this function serves as a comprehensive descriptor of
the joint energy and momentum distribution of the two holes, and
therefore plays a crucial role in characterizing the spectral aspects
of the semi-inclusive two-nucleon emission process induced by
neutrinos. Additionally, the integrated 2h spectral function possesses
an analytical formula in the RFG, providing an added utility to the
factorized approximation. Next we will detail the demonstration of this
analytical formula.

To calculate the integrated spectral function, we follow an
alternative method to the one used above, where we changed to the
CM system of the two holes. We are
 motivated by the fact that the one-body response function of the RFG, 
written in the way (except for a constant factor)
\begin{equation}
R(\omega,q) = 
\int d^3 h_1 d^3h_2 \frac{(m_N^*)^2}{E_1 E_2}  
\theta(k_F-h_1) \theta(h_2-k_F)
\delta(E_2-E_1-\omega)
\delta(\nh_2-\nh_1-q),
\label{respuesta}
\end{equation}
transform into the function \( G(E,H) \), Eq. (\ref{g1}),
if we identify the four-momentum transfer $(\omega,\nq)$ with the
four-momentum of the nucleon pair $(E,\nH)$, and change
\begin{equation}
E_1 \longrightarrow -E_1, \kern 1cm
\nh_1 \longrightarrow -\nh_1, \kern 1cm
\theta(h_2-k_F) \longrightarrow \theta(k_F-h_2).
\end{equation}
In this sense, the function \( G \) could be
formally seen as a kind of analytic
continuation of the one-body response function to the time-like
channel  $Q^2=\omega^2-q^2>0$.

By following the same notation
employed, for example, in \cite{Ama20} for the response function of
the RFG, we will arrive at a very similar result formally for the
integrated 2h spectral function.
We start with Eq. (\ref{g1}).
We first
integrate over $\nh_2$ using the momentum delta function. 
\begin{equation}
G(E,H) = 
\int d^3 h_1 \frac{(m_N^*)^2}{E_1 E_2}  
\theta(k_F-h_1) \theta(k_F-h_2)
\delta(E_1+E_2-E)
\label{g2}
\end{equation}
with $\nh_2=\nH-\nh_1$ and
\begin{equation}
E_2^2 = (m_N^*)^2+(\nH-\nh_1)^2
=E_1^2+H^2- 2 H h_1 \cos\theta_1.
\end{equation}
Hence  for $E_1$ fixed, the values of $E_2$ 
are in the interval $E_{H-h_1} \leq E_2 \leq E_{H+h_1}$, with
\begin{equation}
E_{H-h_1} = \sqrt{(m_N^*)^2+(H-h_1)^2}, \kern 1cm
E_{H+h_1} = \sqrt{(m_N^*)^2+(H+h_1)^2}.
\end{equation}
Since the function $G(E,H)$ only depends on the modulus of $\nH$, we
choose $\nH$ in the $z$-axis and change from spherical coordinates
$(h_1,\theta_1,\phi_1)$ to energy coordinates $(E_1,E_2,\phi)$.  
The volume element transforms as  \cite{Ama20}
\begin{equation}
d^3 h_1= h_1^2 dh_1 d \cos \theta_1 d\phi =\frac{E_1 E_2}{H} dE_1 dE_2 d\phi.
\label{d3h1}
\end{equation}
Then we can write the integral (\ref{g2}) as
\begin{equation}
G(E,H) = \frac{2\pi (m_N^*)^2}{H} \int_{m_N^*}^{E_F} dE_1
\int_{E_{H-h_1}}^{E_{H+h_1}} dE_2 \delta(E_1+E_2-E) \theta(E_F-E_2).
\label{g3}
\end{equation}
Integrating over $E_2$
using the energy delta function, we have $E_2 = E - E_1$, and
\begin{equation}
G(E,H) = \frac{2 \pi (m_N^*)^2}{H}
\int_{m_N^*}^{E_F} dE_1   
\theta(E-E_1 - E_{H-h_1})  
\theta(E_{H+h_1}-E+E_1)
\theta(E_F-E+E_1)
\label{g4}
\end{equation}
Following the notation of Ref. \cite{Ama20}
 we define the dimensionless variables:
\begin{eqnarray}
\lambda = \frac{E}{2m_N^*}, &
\displaystyle \kappa = \frac{H}{2m_N^*}, &
\tau = \kappa^2-\lambda^2= \frac{H^2-E^2}{4(m_N^*)^2}, 
\label{adimen1}\\
\epsilon = \frac{E_1}{m_N^*}, &
\displaystyle \epsilon_F = \frac{E_F}{m_N^*},&
\eta = \frac{h_1}{m_N^*}. 
\label{adimen2}
\end{eqnarray}
In term of these variables that the integral (\ref{g4}) can be written
\begin{equation} \label{g5}
G(E,H)=
\frac{\pi (m_N^*)^2}{\kappa} 
\theta(\lambda-1)\theta(\epsilon_F-\lambda)\theta(-1-\tau)
 \int_{\epsilon_A}^{\epsilon_B} d\epsilon
\end{equation}
The step functions are introduced because $2m_N^* \leq E_1+E_2=E
\leq 2E_F$ and therefore $1<\lambda<\epsilon_F$. On the other hand,
$E^2-H^2>(2m_N^*)^2$ implies $\tau<-1$.

 The integration limits of the integral (\ref{g5}) are 
obtained in Appendix \ref{appb}, and are given by
\begin{eqnarray}
\epsilon_A&=&{\rm Max} 
\left\{\lambda- \kappa \sqrt{1+\frac{1}{\tau}} ,\, 2\lambda - \epsilon_F,\, 1
\right\}
\label{epsilona}\\
\epsilon_B&=&{\rm Min} 
\left\{ \lambda+\kappa \sqrt{1 + \frac{1}{\tau}} ,\, \epsilon_F \right\}
\end{eqnarray}
Finally, we can write the integrated 2h spectral function as
\begin{equation} 
G(E,H)=
\frac{\pi (m_N^*)^2}{\kappa}
\theta(\lambda-1) 
\theta(\epsilon_F-\lambda)
\theta(-1-\tau) 
\theta(\epsilon_B-\epsilon_A)
 (\epsilon_B-\epsilon_A). 
\label{ganal}
\end{equation}
This simple and compact expression for the integrated 2h spectral
function of the RMF is the main result of this section. This can be
considered a universal function, similar to the Lindhard function,
which provides the spectral distribution in the emission of two
particles simply by kinematic considerations, that is, the
phase space. Additionally, it is relativistic and contains the effect
of interaction with the mean field through the effective mass. The
particular case of the RFG is obtained by taking $m_N^*=m_N$. According to
Eq. (\ref{exact}), the semi-inclusive hadronic tensor
 is equal to the function
$G(E,H)$  multiplied by the averaged elementary 2p2h tensor. If this tensor is
slowly varying, it is expected that the cross-section globally follows
the distribution marked by $G$, with small modifications due to the
hadronic tensor. This is seen more explicitly in the factorized
approximation of the cross-section 
\begin{equation}
 \frac{d\sigma_{N_1N_2} }{d T_\mu d \Omega_\mu d^3p'_1 d^3p'_2}
   =
\frac{V}{(2\pi)^9}
\frac{(m_N^*)^2}{E'_1 E'_2 }
\theta(p'_1-k_F)\theta(p'_2-k_F)  
 \sigma_0 L_{\mu \nu} 
w^{\mu \nu} (\np'_1,\np'_2,\langle\nh_1\rangle, \langle\nh_2\rangle) 
G(E'_1+E'_2-\omega,|\np'_1+\np'_2-\nq|).
\label{dsigma}
\end{equation}
In the next section we present results for several observables
obtained from the semi-inclusive 2p2h cross section, and for the
integrated spectral function, using this formalism.

\section{results}

We present numerical predictions for the semi-inclusive
two-nucleon emission reaction off \(^{12}\)C induced by neutrinos and
antineutrinos. The model parameters employed in our calculations
include \(k_F = 225\) MeV/c and \(M^* = m^*_N/m_N = 0.8\), which were
determined in previous studies \cite{Ama18,Mar21a} based on the
investigation of the scaling properties of the $(e,e')$
cross-section. Subsequently, \(m^*_N = 750\) MeV is used as the
effective nucleon mass. The vector energy is $E_v=141$ MeV.

With the same model, we have previously presented results for the
inclusive 2p2h reaction in references
\cite{Mar21a,Mar21b,Mar23a,Mar23b}, which are consistent with the
findings presented here. Additionally, in \cite{Mar24}, we provided
predictions using this model for semi-inclusive 2p2h cross-sections
integrated over four or five variables, complementing those results
with additional observables in this work. Our calculations primarily
utilize the factorized model, and we assess its validity by comparing
it with exact calculations. We also examine the more simplified case
of the phase space model, similar to the one used in Monte Carlo generators,
where it is assumed that the two-particle distribution does not depend
on the elementary hadronic tensor.

\begin{figure}
\includegraphics[width=10cm,bb=95 471 511 765]{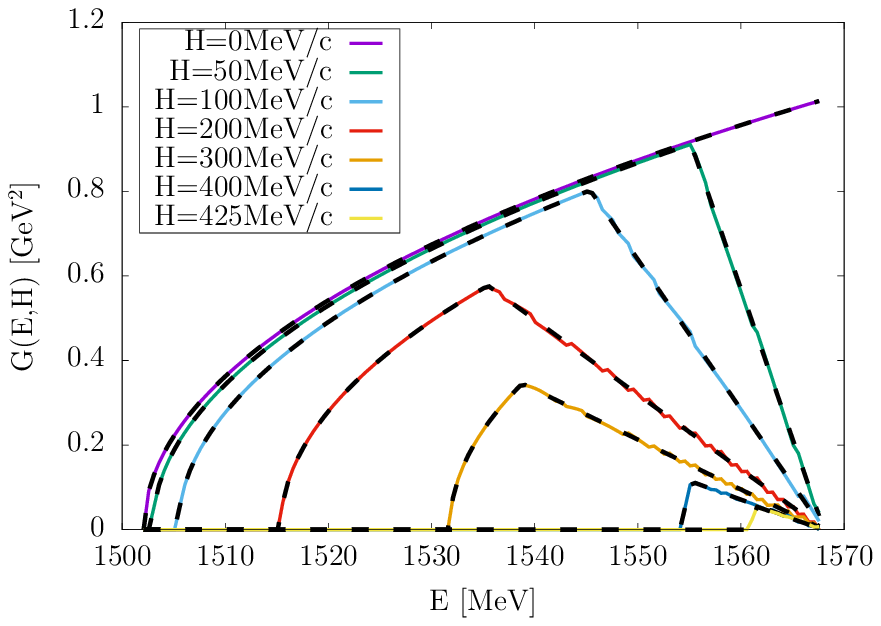}
\caption{ The integrated 2h spectral function $G(E, H)$ 
for various values of momentum $H$ as a
  function of energy $E$. The Fermi momentum is $k_F=225$ MeV/c. Each curve corresponds to a specific
  value of momentum $H$. The dashed lines represent
  the exact analytical result. The solid lines are numerical calculation 
 in the center-of-mass  system.}
\label{fig-g1}
\end{figure}

\begin{figure}
\includegraphics[width=12cm,bb=0 0 360 252]{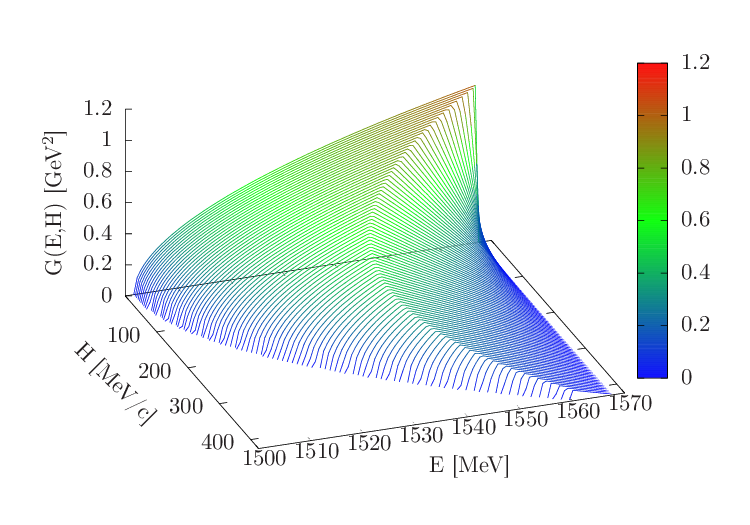}
\caption{ Three-dimensional representation of the integrated 2h
  spectral function $ G(E, H)$. of Fig. 2 }
\label{fig-g2}
\end{figure}

\subsection{Integrated 2h spectral function}

In Fig. \ref{fig-g1}, we present the integrated spectral function
$G(H,E)$ calculated using the analytical formula from
Eq. (\ref{ganal}). The results are compared to a numerical calculation
in the center-of-mass (CM) frame of the two particles. $G(E,H)$ is
plotted as a function of the total 2h energy $E$ for various values of
$H$ ranging from zero up to close to $2k_F$. Note that the accessible values of
$E$ lie between $E_{\min}=2m_N^*$ and $E_{\rm max}=2E_F$, with
$E_F=\sqrt{k_F^2+(m_N^*)^2}=783$ MeV. For $H=0$, all values of $E$ are
allowed, and $G(E,H)$ increases continuously from zero at $E_{\min}$
to its maximum value. Indeed, for $H=0$, nucleon pairs moving
back-to-back in the laboratory frame with any energy can
contribute. As $H$ increases, the value of $E_{\min}$ also increases,
and below this value, $G$ is zero. This means that if the nucleon pair
doesn't have a certain energy, it's not possible for their momenta to
sum up to $H$. As $H$ approaches $2k_F$, the spectral function is
nonzero only when nucleons have energy close to the Fermi energy. For
intermediate values of $H$, the function $G(E,H)$ smoothly increases
with energy until it reaches a point where its derivative is
discontinuous. After this point, it decreases linearly until it
reaches zero at $E=2E_F$. The discontinuity in the derivative is due
to Pauli blocking when the value of $\epsilon_A$ changes abruptly in
Eq. (77).

In Fig. 3, a three-dimensional plot of $G(E,H)$ is presented,
revealing the characteristic structure of this universal function for
the Fermi gas. The function is nonzero only in regions allowed by
kinematics, that is, in the phase space allowed for two holes with
momentum $H$ and energy $E$. It is expected that in a more realistic
model of a finite nucleus, this function would be modified and exhibit
signs of a shell structure, while the maximum momentum would extend
beyond $2k_F=450$ MeV. However, this structure is averaged out and smeared in
neutrino experiments where energy transfer cannot be measured and when
integrating over some of the variables of the final particles. On the
other hand, the factorized approximation allows for modifying the
function $G$ by replacing it with a more realistic function calculated
by other methods. This is another advantage of the factorized model.

\begin{figure}
\includegraphics[width=10cm, bb=140 430 540 760]{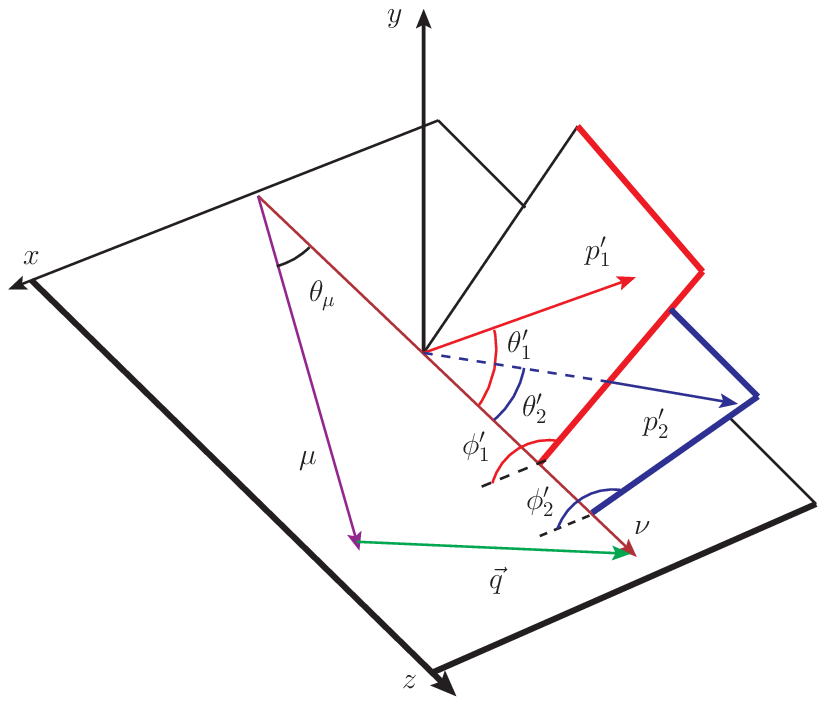}
\caption{ Coordinate system and kinematics in the semi-inclusive
  two-nucleon emission reaction.  }
\label{fig-kin}
\end{figure}

\subsection{Semi-inclusive 2p2h cross section integrated over one energy}

The coordinate system and kinematics for the description of
semi-inclusive 2p2h is shown in Fig. \ref{fig-kin}. We choose the
$z$-axis in the direction of the incident neutrino. 
The neutrino and the final muon directions define the 
 scattering plane $(x,z)$
The directions of the two final
momenta $\np'_i$ and the $z$ axis define the two reaction
planes that form angles $\phi'_1$ and $\phi'_2$, respectively, with
the scattering plane. The angles between $\np'_i$ and the $z$ axis are
$\theta'_i$.

\begin{figure}[ph]
\centering
\includegraphics[width=17cm, bb=58 160 540 730]{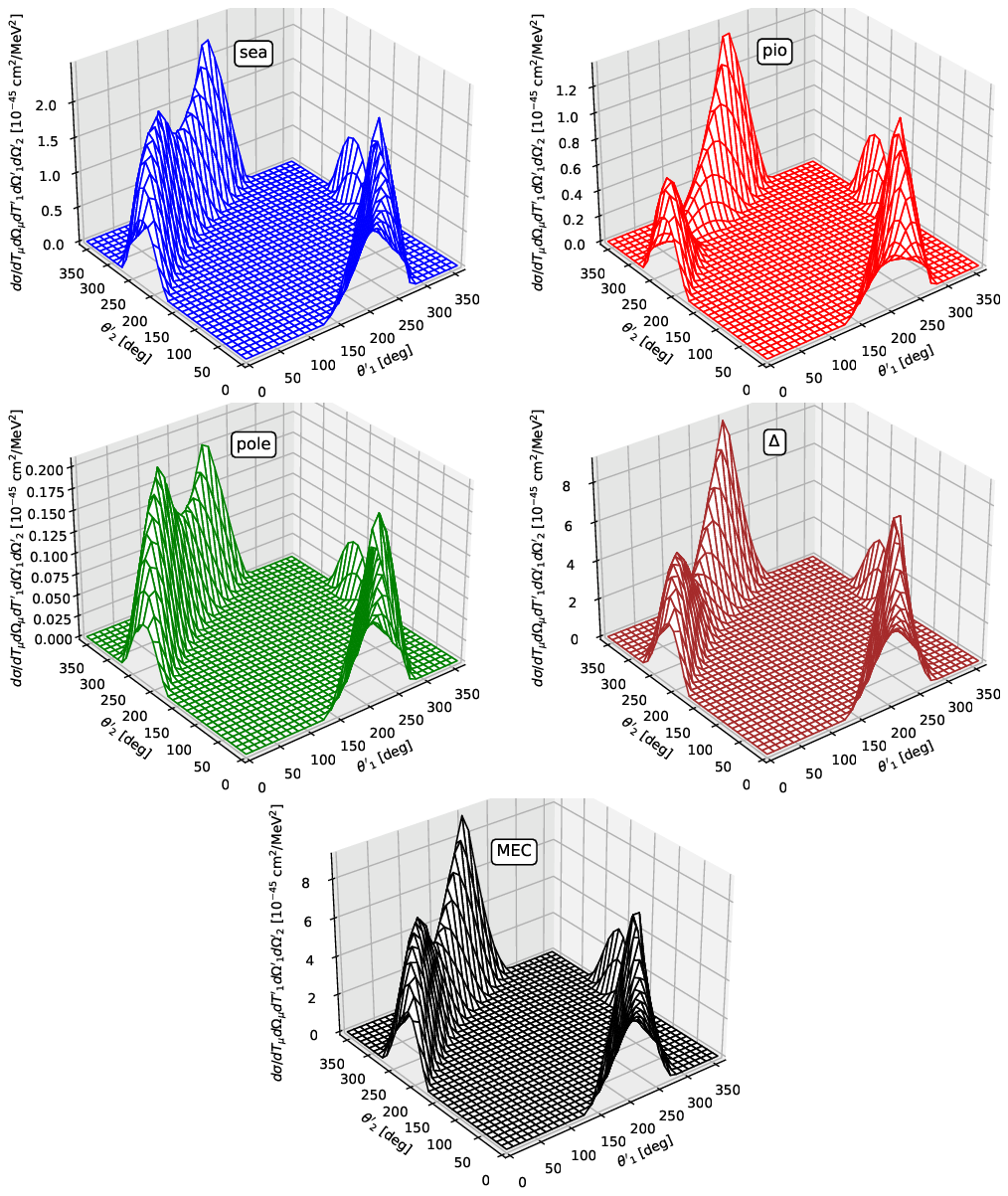}
\caption{The sum of semi-inclusive $^{12}$C$(\nu_\mu,\mu^-pp)$ plus
  $^{12}$C$(\nu_\mu,\mu^-pn)$ cross sections computed with the
  factorized model for in-plane kinematics of the two final particles.
  Incident neutrino energy is $E_\nu= 750$ MeV, muon energy $E_\mu=
  550$ MeV, muon angle $\theta_\mu=15\grado$ and kinetic energy of the
  first particle (a proton) fixed to $T'_1=50$ MeV.  The energy of the
  second nucleon is integrated. In each panel one of the seagull,
  pionic, pole, $\Delta$, and total MEC contributions, are shown.  }
\label{fig-rycke}
\end{figure}

 \begin{figure}[hp]
\centering
\includegraphics[width=16cm, bb=60 200 480 720 ]{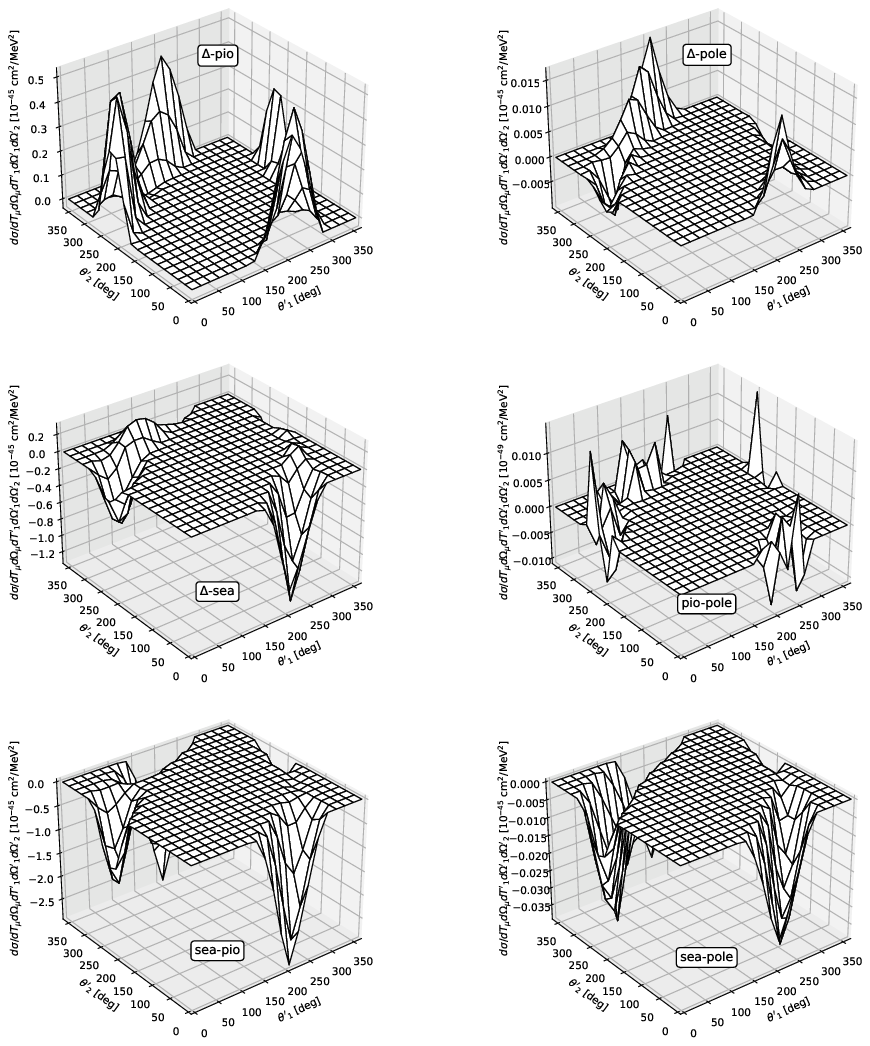}
\caption{The same as Fig. \ref{fig-rycke} for the interferences
  between the different MEC contributions.  Each panel of this figure
  show a different interference pattern corresponding to
  $\Delta$-pionic, $\Delta$-pole, $\Delta$-seagull, pionic-pole,
  seagull-pionic and seagull-pole.}
\label{fig-inter}
\end{figure}

In Figs. \ref{fig-rycke} and \ref{fig-inter}, 
 results for the semi-inclusive 2p2h cross
section for neutrino scattering are presented, integrated over the
energy of the second nucleon and summed over pp and pn pairs.
\begin{equation}
\frac{d\sigma}{dT_\mu d\Omega_\mu dT'_1d\Omega'_1d\Omega'_2}
=\frac{d\sigma_{pp}}{dT_\mu d\Omega_\mu dT'_1d\Omega'_1d\Omega'_2}+
\frac{d\sigma_{pn}}{dT_\mu d\Omega_\mu dT'_1d\Omega'_1d\Omega'_2}.
\end{equation}
Thus particle 1 corresponds to a proton, and particle 2 can be a
proton or neutron. The neutrino energy is fixed at $E_\nu=750$ MeV,
muon angle $\theta_\mu=15$ degrees, and muon energy is 550 MeV. The
kinetic energy of the first proton is fixed at $T'_1=50$ MeV. The
kinematics are coplanar, meaning that the two nucleons exit in the
scattering plane. In Fig. \ref{fig-rycke}, the cross section is
plotted as a function of the angles of the two particles
$(\theta'_1,\theta'_2)$, both ranging from 0 to $2\pi$. This means
that we are simultaneously plotting the 4 cases
$\phi'_1,\phi'_2=0,\pi$. When $\theta'_i>\pi$, the corresponding angle
$\phi'_i=\pi$. This case has been chosen explicitly to allow for
comparison with the calculation of Van Cuyck {\em et al.}
\cite{Cuy17} in the shell model, which is the only available
calculation of this reaction in $^{12}$C.

In Fig.  \ref{fig-rycke} we show the separate contributions of each
of the currents: seagull, pionic, pion pole, $\Delta$, and the total,
while in Fig. and \ref{fig-inter} we present the interferences between
pairs of currents. 
 In Ref. \cite{Cuy17}, the contribution of the
$\Delta$ was not computed. Comparing with Figure 4 of Ref. \cite{Cuy17},
we see that the agreement with the shell model is quite
good, considering that we are using a Fermi gas and that the momentum
transfer is relatively low ($q=265$ MeV/c) for this kinematics, while
$\omega=200$ MeV. Since the quasielastic peak for this value of $q$ is
approximately $\omega_{QE}=q^2/2m_N \simeq 37$ MeV,  we are
in the energy transfer region well above the quasielastic peak, close
to the photon point, where the 2p2h MEC contribution is most
important.

Comparing the magnitude of the separate cross sections with Fig. 4 of
Ref. \cite{Cuy17}, we note that in the case of the seagull and pionic
currents, our cross section is somewhat larger than in the
calculations of Ref. \cite{Cuy17}.  Specifically, the maxima of the
seagull, pion in flight and pion pole cross section are
$(\sigma_S,\sigma_\pi,\sigma_P)$ $\sim (2.2,1.2,0.2) u$ in the RMF,
close to the values $\sim(1.8,1,0.25)u$ obtained in the shell model,
in units of $u=10^{-45}\,{\rm cm^2/MeV^2}$.

\clearpage

The structure of the two peaks observed in Fig. 5 is also similar to
that of the shell model, with approximately the same angular
positions, although in the shell model, they are apparently somewhat
wider. This can be understood given that, in the finite nucleus model,
the momenta are extended and are not limited above the Fermi
momentum. In our case we used the factorized formula (\ref{dsigma}). 
This indicates that the integrated 2h spectral function captures well
the momentum and energy dependence of the semi-inclusive cross section
in more realistic models. Additionally, the averaged values of the
hole momenta in the elementary 2p2h tensor are approximately suitable.

One reason for the agreement with the shell model is that the process
is semi-inclusive and we are integrating over the energy of the second
particle. In the shell model, a sum over occupied shells has been
performed. The integral over energy and sum over shells produce an
smearing of the 2h spectral distribution. Similarly, in the RMF, we
are integrating over holes, producing a similar effect of smearing.

Another reason for the good agreement with the shell model is that we
have performed the correct energy-momentum balance in the kinematics.
This includes taking into account that the given kinetic energy
\(T'_1\) in the semi-inclusive process is the asymptotic energy when
the nucleon is detected. In the shell model, it is the total final
energy of the particle when it is far from the nucleus, where the
nuclear potential is zero. In the case of the RMF the total energy
must include the vector energy. Asymptotically, this must be equal to
the nucleon mass plus the asymptotic kinetic energy. Thus, the correct
energy balance for the first particle is
\begin{equation}
E'_1 + E_v = m_N + T'_1 \Longrightarrow E'_1=m_N+T'_1-E_v.  \label{balance}
\end{equation}
Taking into account that $E_v=141$ MeV, this gives $E'_1=848$ MeV.
From Eq. (\ref{balance}) it follows that the momentum of the particle
in the RMF must be computed as
\begin{eqnarray}
(p'_1)^2= (m_N+T'_1-E_v)^2-(m_N^*)^2  \label{momentum1}
\end{eqnarray}
that gives $p'_1=393$ MeV/c.
Or, assuming non relativistic kinematics for the case of Fig. 5,
\begin{eqnarray}
m_N^* + \frac{(p'_1)^2}{2m_N^*} &=& m_N+T'_1-E_v \Longrightarrow
\nonumber\\ 
(p'_1)^2&=& 2m_N^*(m_N-m_N^*+T'_1-E_v)= 2m_N^*(E_s-E_v+T'_1), 
\label{momentum2}
\end{eqnarray}
with $E_s=m_N-m_N^*=188$ MeV is the (positive) scalar
energy, and $E_s-E_v=47$ MeV \cite{Mar21a}. 
This gives a momentum $p'_1=381$ MeV/c.

Therefore, the differential cross section must be transformed 
with appropriate Jacobian. From Eq. (\ref{momentum2}), we have
$p'_1dp'_1= m_N^*dT'_1$. Hence
\begin{eqnarray}
d^3p'_1= m_N^* p'_1 dT'_1 d\Omega'_1
\end{eqnarray}
and the differential cross section transforms as
\begin{equation}
\frac{d\sigma}{dT_\mu d\Omega_\mu dT'_1d\Omega'_1d\Omega'_2}
=m_N^*p'_1\frac{d\sigma}{dT_\mu d\Omega_\mu d^3p'_1d\Omega'_2}.
\end{equation}
A similar transformation can be obtained for the relativistic case of
Eq. (\ref{momentum1}):
\begin{eqnarray}
d^3p'_1= E'_1 p'_1 dT'_1 d\Omega'_1
\end{eqnarray}
and the differential cross section transforms as
\begin{equation}
\frac{d\sigma}{dT_\mu d\Omega_\mu dT'_1d\Omega'_1d\Omega'_2}
=E'_1p'_1\frac{d\sigma}{dT_\mu d\Omega_\mu d^3p'_1d\Omega'_2}.
\end{equation}

\begin{figure}
\centering
\includegraphics[width=15cm, bb=70 380 500 730]{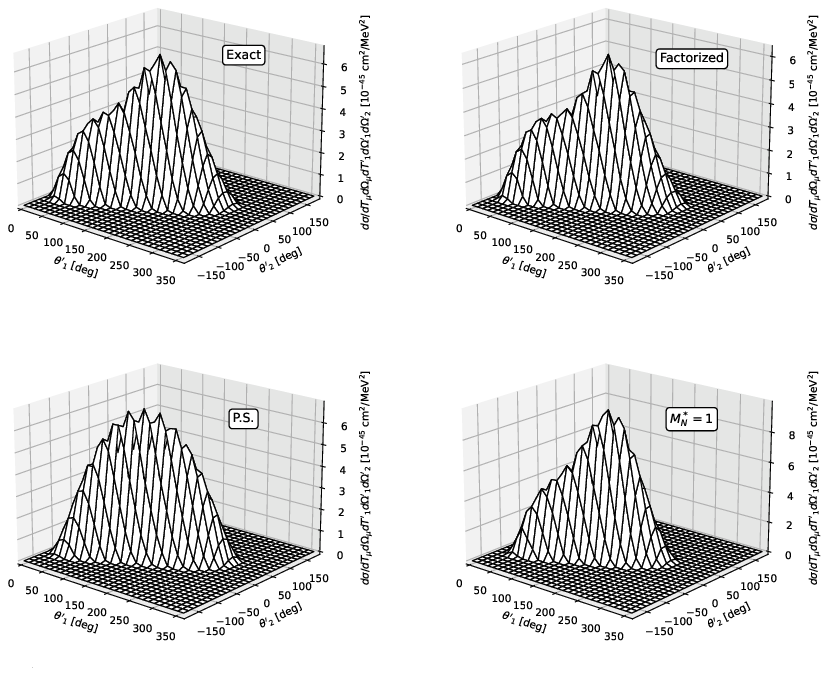}
\caption{
Similar as Fig. \ref{fig-rycke} but for $p'_1=278$ MeV/c, and 
changing the interval of
  the $\theta'_2$ axis from $[0,360]$ to $[-180,180]$.  In each panel
  we show the cross section computed using a different reaction model:
  The exact RMF model, the factorized model, the pure phase-space
  model (P.S.), and the RFG model with a separation energy of 40 MeV.}
\label{fig_models}
\end{figure}

\begin{figure}
\centering
\includegraphics[width=15cm,bb=64 452 525 727]{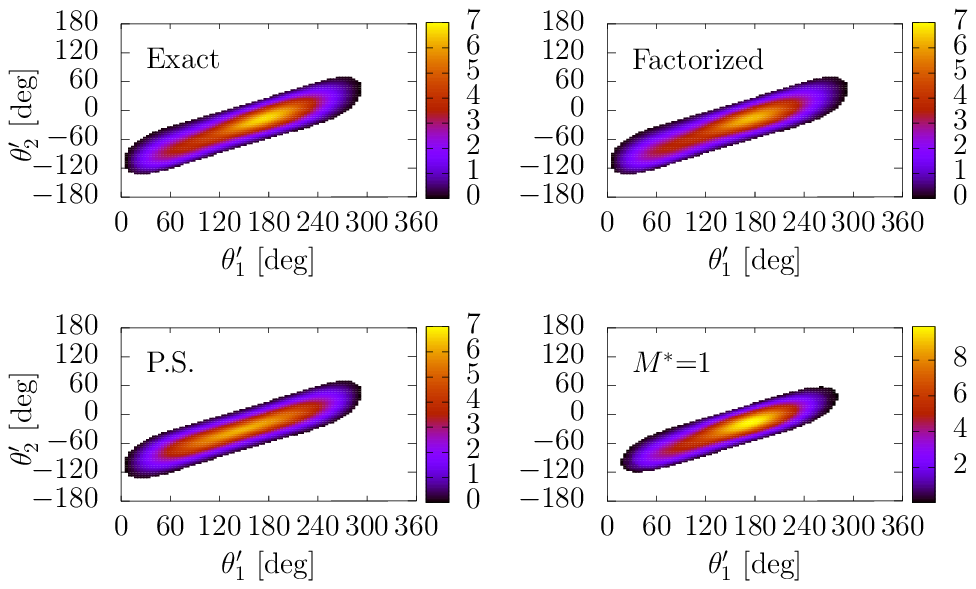}
\caption{ Color map representation of Fig.\ref{fig_models}, where the
  semi-inclusive cross section $d\sigma/dT_\mu d\Omega_\mu dT'_1
  d\Omega'_1 d\Omega'_2$ for the sum of $pp$ and $pn$ neutrino
  emission channels is plotted in units of $10^{-45}$ cm$^2$/MeV$^2$.
}
\label{fig_color}
\end{figure}

The meticulous consideration of the correct energy-momentum balance and the
appropriate transformation of the cross section, accounting for the
asymptotic kinetic energy and the Jacobian factor, ensures the
consistency and reliability of our results. This level of attention to
the details of the theoretical framework is crucial for obtaining
meaningful comparisons with other models, especially when dealing with
different formalisms or experimental measurements.

The comparison between the exact semi-inclusive cross section and the
factorized model is crucial for validating the latter. In Figs. 7 and
8, we present the results for the same cross section as in Figs. 5 and
6, focusing on the same kinematics, but displaying the range of the
angle $\theta'_2$ varying between $-180$ and $180$ degrees for better
visualization, Now the cross section appears as a single peak as a
function of the angles. The top panels in Figs. 7 and 8 depict the
results of both the exact and factorized calculations, revealing
striking similarities. The shape resembles an asymmetric peaked
structure with a shoulder, slightly more pronounced in the factorized
case. Besides this, there are no significant differences, and the
magnitude is the same.

This contrasts with the results obtained using a pure-phase space (PS)
model, shown in the bottom left panels of figs 7 and 8. The pure-phase
space model assumes a constant hadronic tensor, making the results to
follow the shape of the integrated 2h spectral function and normalized
to the inclusive 2p2h cross section. This is analogous to the
procedure employed in Monte Carlo event generators. The PS model
exhibits a peak, but the position of the maximum is shifted, and the
shape of the peak is more symmetric, lacking the shoulder observed in
models with a hadronic tensor. This underscores the importance of
considering the effect of the hadronic tensor in such
reactions. Finally, in the bottom right panel, we compare the
calculation with the RFG without effective
mass, but with a separation energy, 
revealing a significant difference in magnitude compared to the
RMF. Additionally, the peak is somewhat narrower in the RFG case. 
This emphasizes the impact of the effective mass and the necessity of
considering it in the formalism.

\begin{figure}
\includegraphics[width=13.5cm,bb=21 110 575 765]{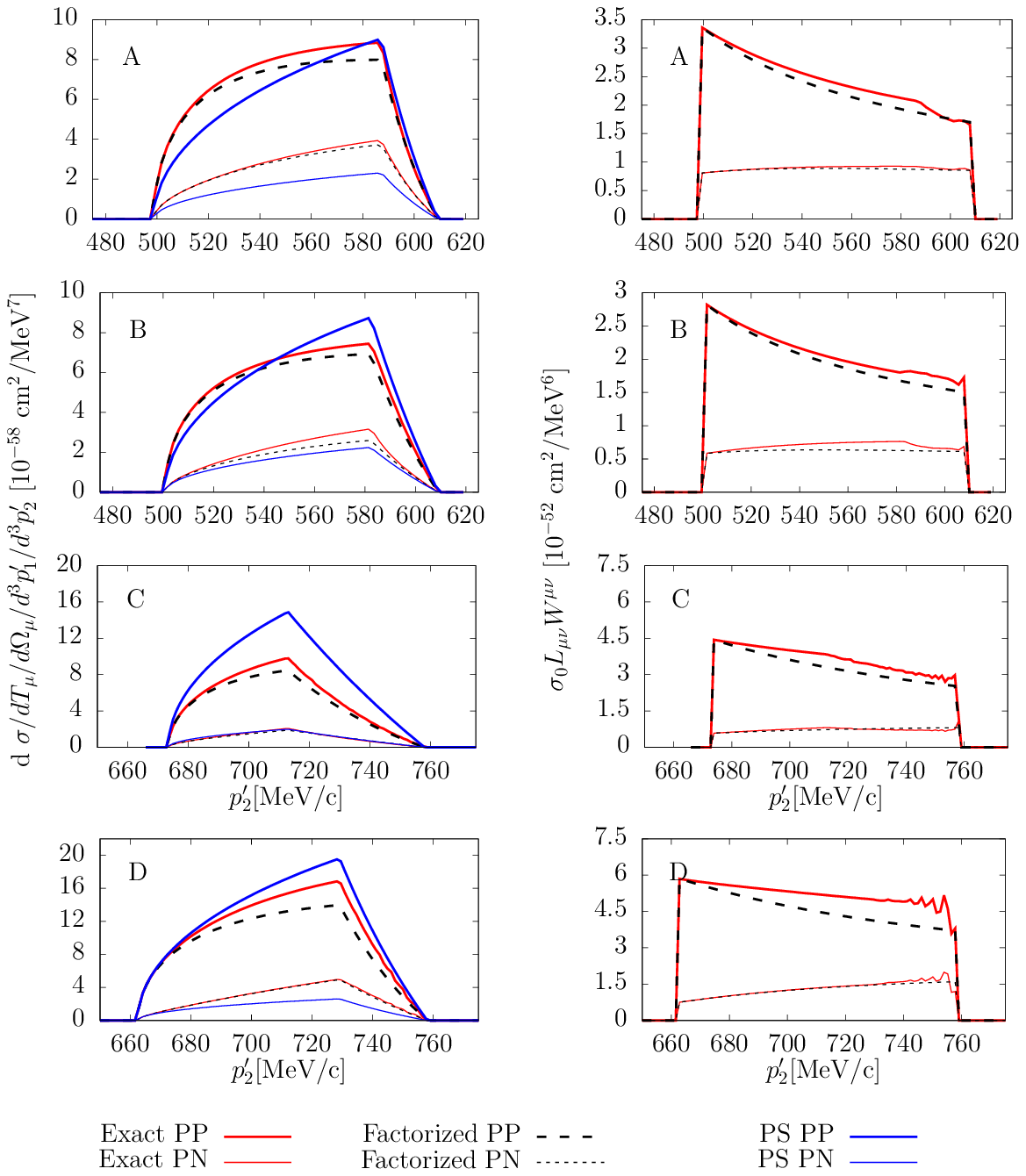}
\caption{Left panels: semi-inclusive pp and pn emission cross section
  for neutrino scattering as a function of $p'_2$. 
   In each panel A--D the lepton kinematics and angles of the exit
  particles are fixed to the values defined in Table \ref{tab_kin}.
  Right panels: the averaged elementary 2p2h hadronic tensor $\langle
  w^{\mu\nu}_{N_1N_2}(\np'_1,\np'_2,\nq,\omega)\rangle$ defined in
  Eq. (\ref{waveraged}), compared to the elementary 2p2h hadronic
  tensor evaluated for averaged hole momenta
  $w^{\mu\nu}_{N_1N_2}(\np'_1,\np'_2,\langle\nh_1\rangle,\langle\nh_2\rangle)$
  as a function of $p'_2$ for the same kinematics.  }
\label{fig-integrando}
\end{figure}

In Fig. 9, we present another test of the factorized approach compared
with the exact result. In the right panels, we display the full
semi-inclusive 2p2h cross section as a function of $p'_2$ for the same
kinematics as in Fig. 5 (panels A, B). In this case, we have fixed the
four angles $\theta'_1, \theta'_2, \phi'_1, \phi'_2$, corresponding to
two representative points on the plot in Fig. 5, which are shown in
Table 1.  We also show the results for kinematics C and D of Table 1,
corresponding to a different lepton kinematics.  Kinematics C and D
were used in Ref. \cite{Mar24} to compute semi-inclusive two nucleon
emission integrated over four variables.  In Fig. 9 we show the
separate pp and pn emission channels for each one of the three models:
exact, factorized, and phase-space models.

Here, it is evident that the choice of averaged values for the hole
momenta in the factorized model does not differ significantly from the
exact result, where the elementary 2p2h hadronic tensor is integrated
over the holes. Both models exhibit very similar behavior. On the
other hand, in the phase-space model, the elementary 2p2h hadronic
tensor is considered a constant and is normalized to the inclusive
2p2h cross section. This leads to the shapes of the curves in the
phase-space model resembling less closely the exact case, even though
the order of magnitude is appropriate due to normalization.

\begin{table}[h]
\caption{ Kinematics used for the results of Fig. \ref{fig-integrando}
of the semi-inclusive 2p2h cross section.
Kinematics A and B are from Figs. 5 and 6. 
Kinematics C and D where also employed in Ref. \cite{Mar24}
to compute the two-folded semi-inclusive cross section.
\label{tab_kin}}
\begin{tabular*}{\textwidth}{@{\extracolsep{\fill}}cccccccc}
\hline\hline\\
\textbf{Kin.}& 
$E_\nu$ [MeV] & $E_\mu$ [MeV] & $p'_1$ [MeV/c] & 
$\phi_1$& $\phi_2$ & $\theta_1$ [deg.] & $\theta_2$ [deg.]
\\
A & 750 & 550 & 278 & 0       & $\pi$    & 172    & 341     \\ 
B & 750 & 550 & 278 & 0       & $\pi$    & 140     & 330    \\ 
C & 950 & 600 & 400 & $\pi$   & $\pi$   & 250    & 355     \\ 
D & 950 & 600 & 400 & 0       & $\pi$    & 50    & 285     \\ \hline
\hline
\end{tabular*}
\end{table}


In the right column of Fig. 9, we present additional results for the
averaged elementary 2p2h hadronic tensor for the same kinematics,
comparing it to the tensor evaluated over averaged
holes. Specifically, we plot the contraction with the leptonic tensor
\[
\sigma_0 L_{\mu,\nu}\langle
w^{\mu\nu}(\np'_1,\np'_2,\nq,\omega)\rangle,
\] 
and compare it to the
contraction with the elementary 2p2h tensor evaluated over averaged
holes
\[
\sigma_0 L_{\mu,\nu}
w^{\mu\nu}(\np'_1,\np'_2,\langle \nh_1\rangle,\langle\nh_2\rangle).
\] 
 Both pp and pn emission channels are shown in the figure.  The
 agreement between both models is quite good, highlighting that the
 elementary hadronic tensor is not constant but depends on the
 kinematics. This dependence is clearly observed in the figure,
 emphasizing the need to consider the full momentum and energy
 dependence in the tensor, as is done in the factorized model, rather
 than assuming a constant value, as in the phase-space model.

Finally, in Figs. 10 and 11, we explore the semi-inclusive cross
section integrated over one energy, focusing on different increasing
values of the proton momentum $p'_1$. In Fig. 10 the lepton kinematics
is the same as in Fig. 5 and $p'_1=278$, 393, and 600 MeV/c.  In
Fig. 11 the lepton kinematics is different with larger neutrino energy
$E_\nu=950$ MeV, $E_\mu=600$ MeV, and $\cos \theta_\mu=0.85$.  This
correspond to 'Kinematic \#1' from Ref. \cite{Mar23b}, where we
computed the semi-inclusive cross section integrated over four
variables. The three different values of proton momentum in Fig 11 are
$p'_1= 400$ MeV/c, 600 MeV/c, and 800 MeV/c.

An important general feature that emerges from these angular
distribution plots is that the two nucleons tend to be emitted in
opposite directions. The back-to-back tendency is only
approximate. This means that the angle between $\mathbf{p}'_1$ and
$\mathbf{p}'_2$ is greater than 90 degrees and predominantly closer to
180 degrees. For example, in the top panels of Fig. 10, the maximum of
the cross section occurs around $\theta'_1 \sim 200$ degrees and
$\theta'_2 \sim 350$ degrees. That is, $\theta'_2 - \theta'_1 \sim 150$ 
degrees. It can also be seen in the values of Table 1,
corresponding to angular positions where the cross section is
significant, where the differences between the fifth and fourth
columns are $\theta'_2 - \theta'_1 = 169$, 190, 105, and 235
degrees. This approximate tendency of nucleons to be emitted back-to-back will
also be independently confirmed by the results of the next section.

These plots provide insights into how the distributions evolve and
change shape as the energy of the detected nucleon increases. The
strength shifts angularly, and the main peak changes its position. In
Fig. 10, as the energy increases, the shoulder observed in Fig. 7
becomes more pronounced, eventually splitting into two distinct
peaks. However, for $p'_1=600$ MeV/c, there is a return to a single
peak but in a different angular position.  Something similar happens
for the kinematics of Fig. 11.

\begin{figure}
\centering
\includegraphics[width=16cm, bb=60 200 480 720 ]{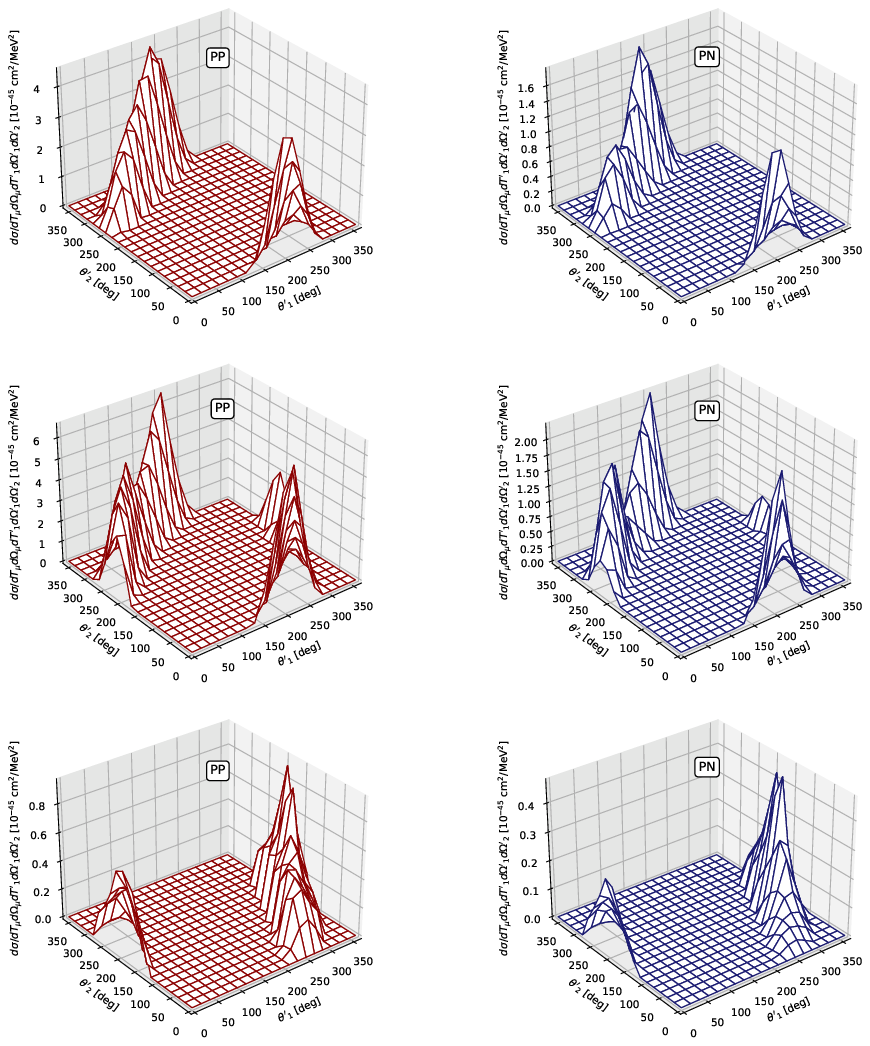}
\caption{Semi-inclusive two-nucleon emission cross sections integrated
  over the second particle, for neutrino scattering 
and separated in proton-proton (pp) and
  neutron-neutron (nn) channels.  The kinematics as in Fig.5 but with
  different values of proton momentum, $p'_1=278$ MeV/c in the
  top panel, 393 MeV/c in the middle panel, and 600 MeV/c in the
  bottom panel. The pp channel is depicted in red on the left side,
  while the pn-channel is represented in blue on the right side.}
\label{fig-channels}
\end{figure}

\begin{figure}
\centering
\includegraphics[width=16cm, bb=60 200 480 720 ]{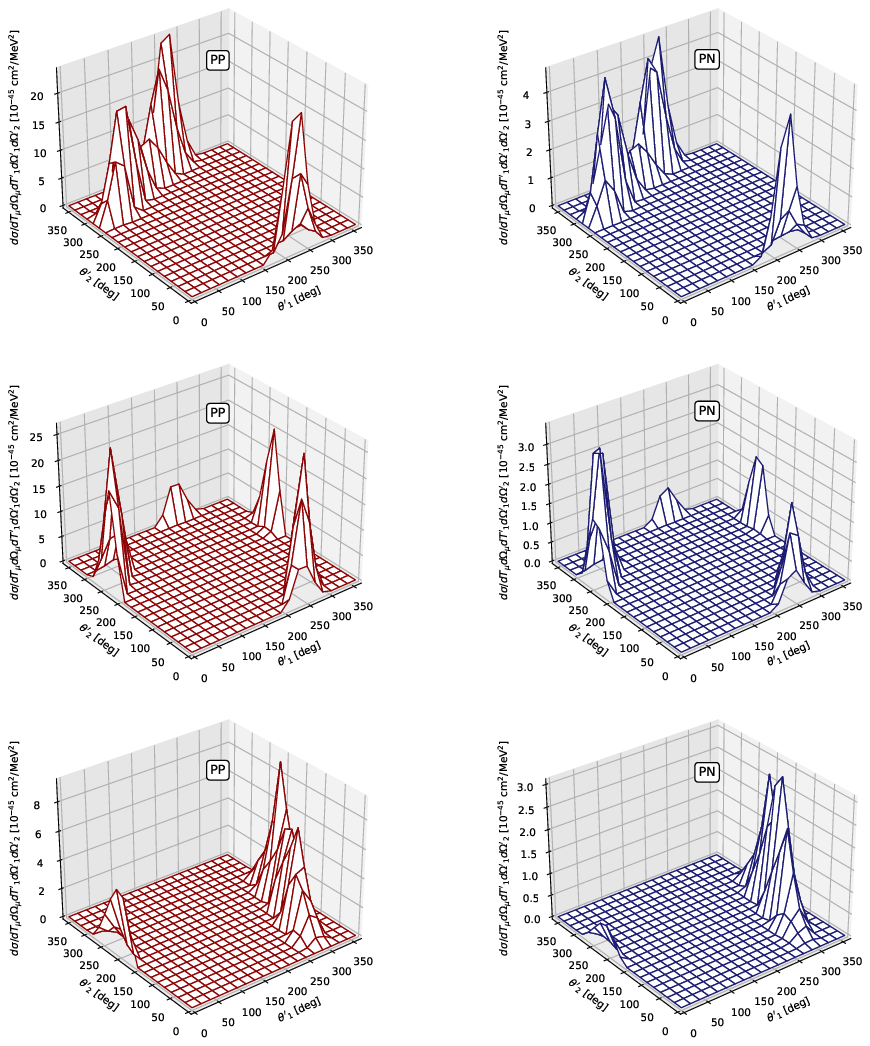}
\caption{ The same as Fig. \ref{fig-channels} for a different 
  kinematics given by
  $E_\nu=950$ MeV $E_\mu=600$ MeV and $\cos  \theta_\mu=0.85$
('Kinematic \#1' from \cite{Mar23b}). The panels
  represent three different values of proton momentum $p'_1= 400$
  MeV/c  (top panels), 600 MeV/c (middle panels), and 800 MeV/c
  (bottom panels).
}
\label{fig_channels2}
\end{figure}

\subsection{Semi-inclusive 2p2h cross section integrated over the muon}

In this last subsection, we explore another observable of interest in
semi-inclusive two-nucleon emission: the cross section integrated over
the muon energy and angles, as well as the final nucleon angles.
\begin{equation} \label{nieves}
\frac{d\sigma_{N_1N_2}}{dp'_1dp'_2}
=2\pi p'_1{}^2 p'_2{}^2
 \int_{T_{\rm min}}^{T_{\rm max}} dT_\mu 
\int_{u_{\rm min}}^1 d\cos\theta_\mu 
\int d\Omega'_1 
\int d\Omega'_2 
\frac{d\sigma_{N_1N_2}}{dT_\mu d\Omega_\mu d^3p'_1d^3p'_2}
\end{equation}
where the integration limits are given below.

The motivation for this study is to compare predictions with the
Valencia model and recent results obtained within the NEUT generator,
as published in the reference \cite{Sob20}.
This comparison is valuable because the Valencia model also employs an
interacting relativistic Fermi gas, introducing interaction through a
different effective interaction. Additionally, it includes effects
such as short-range and long-range correlations of the RPA type, while
neglecting the interference of the direct and exchange current matrix
elements, among other considerations detailed in \cite{Sob20}. On the
other hand, results from the NEUT generator are representative of what
is expected from a model that applies a phase-space approximation for
the 2p2h emission, neglecting the dependence of the hadronic tensor on
the exclusive variables $\np'_1, \np'_2, \nh_1, \nh_2$. In contrast,
we apply the factorized approximation of the RMF model, which has been
tested in the previous subsection and yields results very similar to
the shell model of \cite{Cuy17}.

The factorized approximation in this case is convenient because it
saves us from the computation of an eight-dimensional integral, as
required by the exact calculation. The factorization allows us to use
the analytical formula for the $G(E,H)$ function, introducing the
elementary 2p2h hadronic tensor evaluated at averaged hole
momenta. Thus, we are left with a six-dimensional integral that needs
to be computed numerically.

First we examine the integration limits that we have written
explicitly in Eq. (\ref{nieves}) for the muon kinetic energy 
$T_{\rm   min}<T_\mu<T_{\rm max}$ and angle $u_{\rm min}<
\cos\theta_{\mu}<1$. Note that these integration limits are specific for our
RMF+MEC approach and are not general. 

We maintain a fixed neutrino energy, \(E_\nu\), while considering
\(p'_1\) and \(p'_2\) as the fixed momenta of the emitted
nucleons. Consequently, the energies \(E'_1\) and \(E'_2\) are
also predetermined. The conservation of energy is expressed by the equation
\(E_\mu = E_\nu + E_1 + E_2 - E'_1 - E'_2\).
In our model the initial
hole energies are bounded within the range \(m_N^* < E_i <
E_F\).
This bounding of initial particle energies
inherently limits the energy available for the muon, ensuring
\(E_\mu\) falls within a defined range. 
\begin{equation}
E_\nu + 2m_N^* - E'_1 - E'_2< E_\mu < 
E_\nu + 2 E_F - E'_1 - E'_2. 
\end{equation}
This means that the integration limits must be
\begin{eqnarray}
T_{\rm min} &=& E_\nu-m_\mu + 2m_N^* - E'_1 - E'_2,\\
T_{\rm max} &=& E_\nu-m_\mu + 2E_F - E'_1 - E'_2.
\end{eqnarray}

The lower limit of \(\cos\theta_\mu\) for fixed \(E_\nu\) and \(E_\mu\)
is due to the fact that the 2p2h response can be neglected if the
energy transfer is below the threshold energy to kick two initially
at-rest particles that are emitted with a total momentum \(q\) (frozen
nucleon approximation). Therefore, the dominant contribution to the
integral requires that
\begin{equation}
 E\nu-E\mu = \omega > \sqrt{4(m_N^*)^2+q^2}-2 m_N^*
\end{equation}
From where
\begin{equation} \label{ofrozen}
(E_\nu -E_\mu +2 m_N^*)^2 \ge 4(m_N^*)^2+q^2.
\end{equation}
On the other hand the momentum transfer is given by
\begin{eqnarray}
q^2 &=& (k-k')^2 = k^2+k'^2-2kk'\cos \theta_\mu\\
&=& E_\nu^2 + E_\mu^2-m_\mu^2- 2E_\nu\sqrt{E_\mu^2-m_\mu^2} \cos \theta_\mu
\end{eqnarray}
Substituting this value of the momentum transfer in Eq. (\ref{ofrozen}) and expanding the square,
\begin{eqnarray}
E_\nu^2 +E_\mu^2 - 2E_\nu E_\mu + 4 (m_N^*)^2 + 4 m_N^*(E\nu-E_\mu) 
&&
\nonumber \\ 
&& \kern -5cm
\ge 4(m_N^*)^2+E_\nu^2 + E_\mu^2-m_\mu^2
-2E_\nu\sqrt{E_\mu^2-m_\mu^2} \cos \theta_\mu.
\end{eqnarray}
Solving for $\cos \theta_\mu$ and simplifying we obtain the lower limit:
\begin{equation}
\cos \theta_\mu \ge 
\frac{2 E_\nu E_\mu -m_\mu^2 - 4m_N^* (E_\nu-E_\mu)}{2E_\nu \sqrt{E_\mu^2-m_\mu^2}}
\equiv u_{\rm min}
\end{equation}
Applying these integration limits when performing the numerical
integral helps speed up the calculation, as it avoids calculating the
2p2h hadronic tensor for kinematics that are suppressed by these
limits.

\begin{figure}
\centering
\includegraphics[width=16cm , bb= 80 390 530 780]{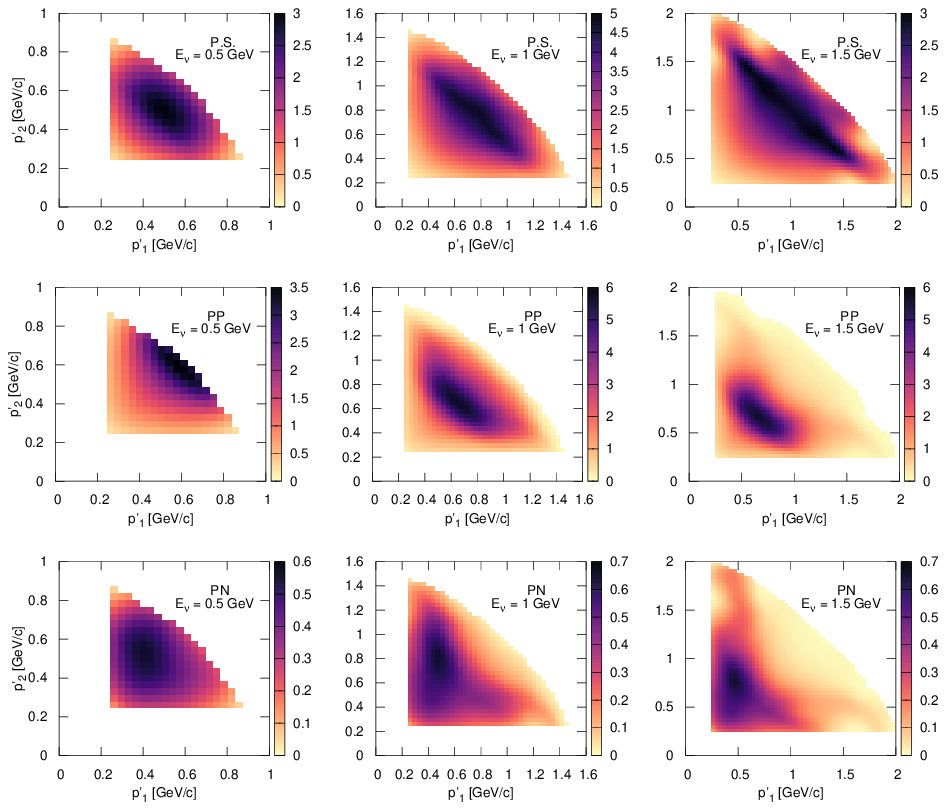}
\caption{ Integrated semi-inclusive cross section
  $d\sigma_{N_1N_2}/dp'_1dp'_2$ as a function of outgoing nucleon
  momenta for three neutrino energies: $E_\nu= 500$ MeV, 1000 MeV, and
  1500 MeV.  In the top panels we show the pure phase-space (P.S.)
  results.  In the middle and bottom panels we show the pp and pn
  emission channels, respectively, computed with the factorized RMF
  model.  The phase space is normalized to the inclusive total
  neutrino cross section.
  The units are $10^{-38}$ cm$^2$/GeV$^2$.}
\label{fig-nieves1}
\end{figure}

\begin{figure}
\centering
\includegraphics[width=16cm , bb= 80 390 530 780]{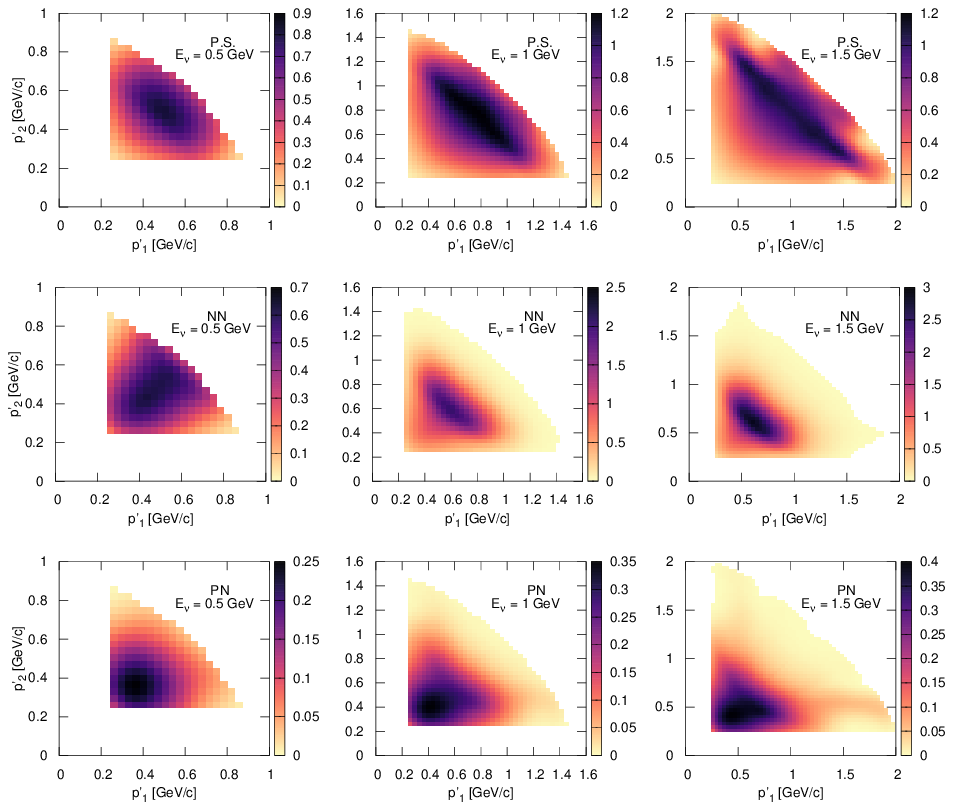}
\caption{
The same as Fig. \ref{fig-nieves1} for antineutrino scattering.
}
\label{fig-nieves2}
\end{figure}

The integrated cross section from Eq. (\ref{nieves}) is shown in
figures \ref{fig-nieves1} for neutrinos and \ref{fig-nieves2} for
antineutrinos. We present the results for three incident neutrino
energies: $E_\nu=500$, 1000, and 1500 MeV. These values are the same
as those used in figures 11 and 12 of ref. \cite{Sob20} for neutrino
scattering, considering the same observable for comparison. In our
results, we employ two models. One is the pure phase space (top panels
of Fig. \ref{fig-nieves1}, \ref{fig-nieves2}), where the elementary
2p2h hadronic tensor is not included. Specifically, in the P.S. model,
we set $L_{\mu\nu}w^{\mu\nu}(\np'_1,\np'_2,\nh_1,\nh_2)=
1$. Therefore, the model only contains the integrated 2h spectral
function, and it is normalized with a constant so that the PS total
cross section coincides with the factorized one. We present these
results as a way to observe the effect of the elementary 2p2h tensor
hadronic in this observable.  The other calculations shown in the
middle and bottom panels of Fig. 11 correspond to the emission
channels of pp and pn, respectively. In these cases, the first
particle, \(p'_1\), is always a proton, while \(p'_2\) can be either a
proton or a neutron.

\begin{figure}
\includegraphics[width=8cm,bb= 129 482 466 722]{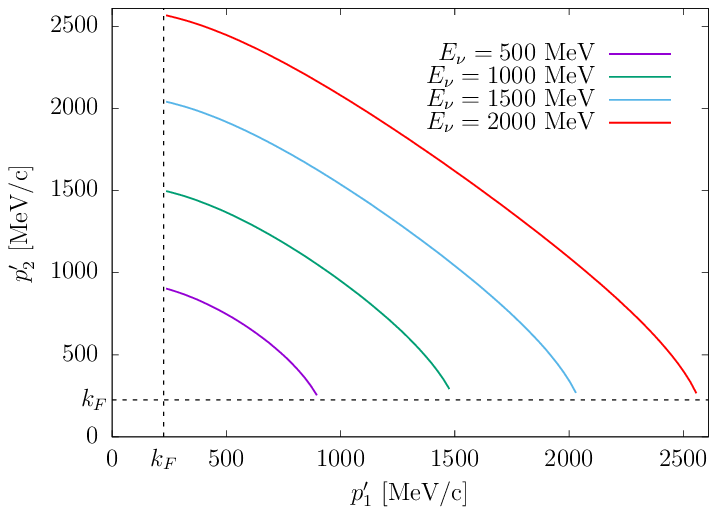}
\caption{Maximum values of outgoing nucleon momentum $p'_2$ as a
  function of $p'_1$ for various neutrino energies}
\label{fig_maximum}
\end{figure}

The first thing we notice is the shape of the distribution in the
plane $(p'_1,p'_2)$. The cross section is zero beyond an a surface
that is approximately a quarter of a circle centered at the point
$(p'_1,p'_2)=(k_F,k_F)$, because $k_F$ is the minimum value of
\(p'_i\).  The boundary of the surface is determined by energy
conservation. The curve defining the boundary of the surface can be
written as a function of \(p'_2\) in terms of \(p'_1\).
In fact,  we use energy conservation
\begin{equation}
E'_1 + E'_2= E_\nu-E_\mu + E_1+E_2  
\end{equation}
and apply that the maximum energy of the holes
is the Fermi energy, and the minimum energy of the muon is the muon
mass: 
\begin{eqnarray}
E_1+E_2 < 2 E_F \\
m_\mu< E_\mu  \Longrightarrow -E_\nu < -m_\mu
\end{eqnarray}
Then we have 
\begin{equation}
E'_1 + E'_2 <  E_\nu-m_\mu+2E_F \Longrightarrow
E'_2 <  E_\nu-m_\mu+2E_F-E'_1.  
\end{equation}
Taking the square of the last inequality and solving for the momentum,
we obtain the limiting curve
\begin{equation}
p'_2 < p'_2{}_{\rm max}= 
\sqrt{\left(E_\nu-m_\mu+2 E_F-\sqrt{p_1^{'2}+(m_N^*)^2}\right)^2 - (m_N^*)^2}
\end{equation}
The curve $ p'_2{}_{\rm max}$ as a function of $p'_1$ is plotted in
Fig. \ref{fig_maximum} for several values of the neutrino
energy. Comparing with Figs. \ref{fig-nieves1} and \ref{fig-nieves2} we
see that they explain the border of the distribution.

The second observation from figs. \ref{fig-nieves1} and
\ref{fig-nieves2} is that the peak of the distribution in the case of
the phase-space model (PS) shifts towards larger momenta as the
neutrino energy increases. However, in the case of the factorized
model, the peak remains more or less in the same position in the
$(p'_1, p'_2)$ plane, both in pp and pn emission. This is due to the
inclusion of the elementary 2p2h hadronic tensor, which has a peak
around the $\Delta$ resonance. The results in the figure show that the
position of this peak does not change much with increasing neutrino
energy.

A possible explanation of the invariance of the position of the distribution
peak is based on the assumption of back-to-back dominance of the
final particles, as we have seen in the angular distributions of the
last section, together with the additional assumption of dominance of the
$\Delta$-forward diagrams of the MEC for pp emission.  In fact the
argument is the following. To simplify this discussion we set the
effective mass equal to the nucleon mass. The assumption is that the
greatest contribution to the cross section comes from back-to-back
nucleons. From momentum conservation this gives
\begin{equation} \label{bb1}
\nq+\nh_1+\nh_2= \np'_1+\np'_2 \simeq 0 \Longrightarrow \nq+\nh_1\simeq -\nh_2.
\end{equation}
Assuming the dominance of the $\Delta$-forward current, 
if the first nucleon absorbs the energy-momentum
transfer (diagram (f) of Fig. 1), then the maximum contribution occurs
when the $\Delta$ propagator is at its maximum, i.e., we are close to
an on-shell $\Delta$:
\begin{equation}
(\omega+E_1)^2-(\nq+\nh_1)^2 \simeq M_\Delta^2.
\end{equation}
 Using the result from the back-to-back condition (\ref{bb1}) we obtain
\begin{equation}
\omega+E_1 \simeq \sqrt{h_2^2+M_\Delta^2}.
\end{equation}
Using this result in the energy conservation 
\begin{equation}
2E'_1 \simeq E'_1+E'_2 = \omega + E_1+E_2 \simeq \sqrt{h_2^2+M_\Delta^2}+E_2.
\end{equation}
Finally we obtain
\begin{equation}
E'_1 \simeq \frac12 \left( \sqrt{h_2^2+M_\Delta^2}+E_2 \right).
\end{equation}
This suggests that the position of the maximum contribution does not
depend strongly on the energy of the incoming neutrino, as long as we
are in the back-to-back configuration and the $\Delta$-forward current
dominates. This could explain the observed behavior in the pp
distributions of Fig. \ref{fig-nieves1}, where the position of the
peak remains relatively stable even with increasing neutrino energy.

If we give values to the hole momentum $h_1=0,k_F$ we obtain
\begin{eqnarray}
h_1=0 
&\Rightarrow & 
E'_1 \simeq \frac12(M_\Delta+m_N)=1086\, {\rm MeV} \Longrightarrow 
p'_1 \simeq 546\, {\rm MeV/c} 
\\
h_1=k_F 
&\Rightarrow & 
E'_1 \simeq \frac12 \left( \sqrt{k_F^2+M_\Delta^2}+E_F \right)=1108 \,{\rm MeV}
\Longrightarrow 
p'_1 \simeq 588 \,{\rm MeV/c} 
\end{eqnarray}
The values obtained under our assumption, \(p'_1=546-588\), are not
very far from the actual position of the peak in Fig. 12, \(p'_1\simeq
600-700\). We attribute the difference to the approximations
made to obtain the rough estimation of the maximum since the nucleons
do not strictly emerge back-to-back. Other contributions in the MEC,
the neglect of the effective mass effect, and additional factors also
contribute to the discrepancy. However, the result is reasonably
sound, allowing us to suggest that the hypothesis of back-to-back
dominance has some relevance to the results in Fig. 12.

The results of Fig. 12 can be compared with those of Figs 11 and 12 of
Ref. \cite{Sob20}, where the same cross sections were computed for the
same kinematics for pp and pn emission with the Valencia model of 2p2h
emission and the NEUT event generator, which includes the final state
interaction (FSI) with an intranuclear cascade model. Note that our results
are directly the results of the primary vertex of the interaction and
do not include FSI, which is expected to change the distribution
slightly and possibly make the peak narrower.

Differences are observed between our results and those of the Valencia
model. For example, the Valencia model predicted a maximum of the pp
distribution for \(p'_1\sim 0.9-1.2\) GeV/c and \(p'_2\sim 0.4-0.5\)
GeV/c, attributed in Ref. \cite{Sob20} to the $\Delta$ current,
whereas in our case, as mentioned above, the peak is observed at
\(p'_1=p'_2\sim 0.6-0.7\) GeV/c.

The comparison with NEUT also does not agree with our pure phase space
calculation, presumably because we do not normalize for each value of
\(q, \omega\), but to the total result of pp + pn emission. In this
way, it is seen that it is important to include at least the inclusive
responses as a function of \(q, \omega\) to obtain some structure
apart from the pure spectral function.

In the case of pn emission, another discrepancy is observed when
comparing with the Valencia model. Both distributions are asymmetric
when changing from proton to neutron. However, in our case, it is
observed that at the maximum, the neutron exits with more energy than
the proton, while the opposite occurs with the Valencia model.

An explanation for our result that the neutron is more energetic
than the proton in pn emission can be made similarly to that given in
Ref. \cite{Sob20}. However, in our case,
where we only include MEC, the same deduction leads us to the
conclusion that the neutron predominantly exits with more energy than
the proton. Since the explanation in Ref. \cite{Sob20} is not
detailed, we cannot draw strong conclusions about the
differences. Therefore, we provide a more in-depth explanation of our
results.

\begin{figure}
\includegraphics[width=11cm,bb= 90 540 510 700]{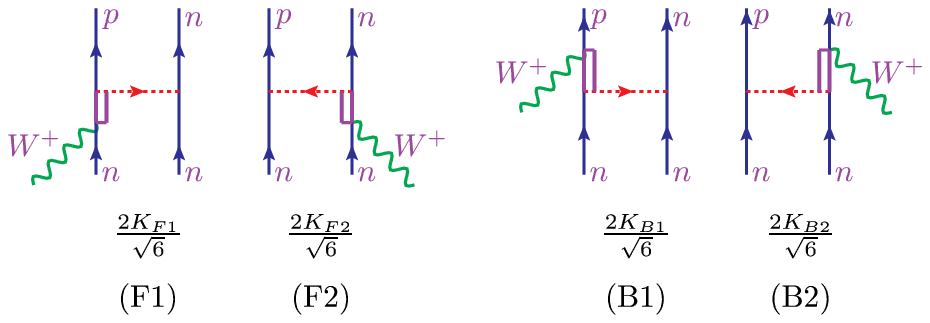}
\caption{Feynman diagrams representing the matrix elements  for 
forward ($K_{F1}$,
  $K_{F2}$) and backward ($K_{B1}$ and $K_{B2}$) $\Delta$ current 
for pn emission.}
\label{fig-kas}
\end{figure}

The argument is based on the assumption that the Delta current is the
main contribution to the cross section. Under this hypothesis we
compute the matrix element of the $\Delta$ current between the initial
nn and final pn pair.
From Eqs. (\ref{deltaF}) and (\ref{deltaB})
the $\Delta$ matrix elements are
\begin{eqnarray}
j^\mu_{\Delta F}
&=&
\langle pn| 
\frac{1}{\sqrt{6}}
\left[ 
2\tau_+^{(2)}K_{F1}^\mu
+2\tau_+^{(1)}K_{F2}^\mu
-I_{V+}(K_{F1}-K_{F2})^\mu
\right]
| nn \rangle,
\label{deltaF}
\\
j^\mu_{\Delta B}
&=&
\langle pn| 
\frac{1}{\sqrt{6}}
\left[ 
2\tau_+^{(2)}K_{B1}^\mu
+2\tau_+^{(1)}K_{B2}^\mu
+I_{V+}(K_{B1}-K_{B2})^\mu
\right]
| nn \rangle,
\label{deltaB}
\end{eqnarray}
 Using the basic matrix elements of the isospin
operators (\ref{isospin}) 
\begin{eqnarray}
\langle pn | I_{V+} | nn \rangle &=& -2, \\
\langle pn | \tau_{+}^{(1)} | nn \rangle &=& 2, \\
\langle pn | \tau_{+}^{(2)} | nn \rangle &=& 0. 
\end{eqnarray}
we obtain
\begin{eqnarray}
j^\mu_{\Delta F}
&=&
\frac{2}{\sqrt{6}}
\left(K_{F1}^\mu+K_{F2}^\mu\right)  \label{forward}
\\
j^\mu_{\Delta B}
&=&
\frac{2}{\sqrt{6}}
\left( 3K_{B2}^\mu-K_{B1}^\mu\right).
\label{backward}
\end{eqnarray}
Here the functions $K_{F1}^\mu$, $K_{F2}^\mu$, $K_{B1}^\mu$ and $K_{B2}^\mu$
correspond to the diagrams of Fig. \ref{fig-kas}.  It is fundamental
to remember that we are considering the case where particle $p'_1$ is
a proton and particle $p'_2$ is a neutron, as specified in
fig. \ref{fig-kas}.  The argument applies similarly when considering a
neutron as particle 1 and a proton as particle 2, obtaining the same
results.

\begin{figure}
\centering
\includegraphics[width=17cm,bb=46 402 553 720 ]{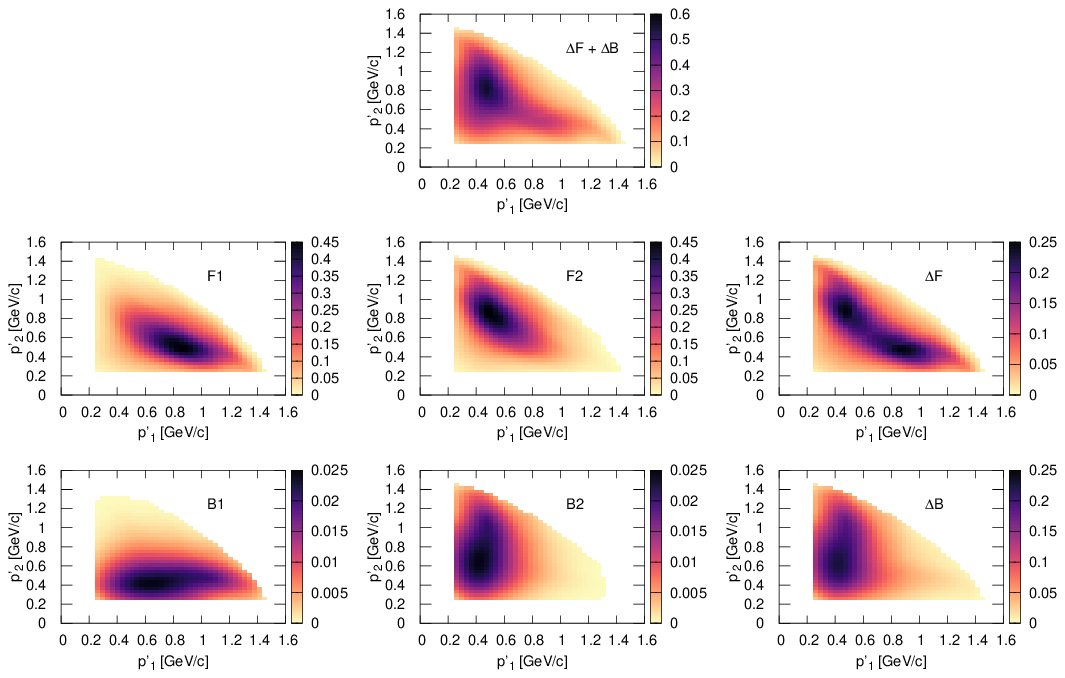}
\caption{ Color map of the cross section for neutrino-induced $pn$
  emission. We display the separate contributions of the forward,
  $K_{F1}$, $K_{F2}$, and backward, $K_{B1}$, $K_{B2}$ terms of the
  $\Delta$ current, as well as the total contribution of the forward
  and backward, and finally the total contribution of the $\Delta$
  current alone. The neutrino energy is $E_\nu=1$ GeV. The units of
  the cross section are the same as in Fig. 12}
\label{fig-final}
\end{figure}

From Fig. \ref{fig-kas} and Eqs. (\ref{forward},\ref{backward}), it is
evident that there are four main contributions to the cross
section. The interaction with the initial nn pair result in a
particle-hole excitation connected to the $W^+$, with the final particle
being a proton (diagrams F1 and B1) or a neutron (diagrams F2 and B2). 
In the case of F1, the final proton  receives a significantly higher
energy-momentum transfer, while in the case of F2, the neutron becomes
much more energetic. 
These two possible contributions have equal
strength. The same can be said for backward diagrams B1 and B2. 

Fig. \ref{fig-final} illustrates the single contributions of $K_{F1}$,
$K_{F2}$, $K_{B1}$ and $K_{B2}$ to the cross section, along with the
total forward and backward contribution.

We observe that the cross section obtained with the term F1 alone
results in a distribution where the proton (particle 1) is more
energetic than the neutron (particle 2) at the maximum. On the
contrary, the contribution of the term F2 is exactly the same as F1,
changing the proton for the neutron. As a result, the combined
contribution of the two terms F1 + F2 produces two maxima
approximately corresponding to the positions of the maxima of F1 and
F2 (note that in the total result there is an interference of F1 with
F2 that is also taken into account).

We now examine the individual contribution of the backward terms B1
and B2. The contribution of the term B2 has been calculated using the
current \(j_B = \frac{2}{\sqrt{6}} K_{B2}\) without the factor of 3 in
Eq. (\ref{backward}), so that both terms enter with the same weight. Then,
  in Fig. \ref{fig-final}, we see again that protons have more energy
  in the distribution B1, while neutrons are the most energetic in the case of
  B2. Again, the two distributions, B1 and B2, are obtained from each other
  by changing the proton for the neutron. However, in the total
  distribution B1 + B2, the term B2 carries a factor of 3 with respect
  to B1. Note that the currents are squared, meaning that the term B2
  contributes actually with a factor of 9 with respect to B1. This
  results in the total backward distribution predominantly showing more
  energetic neutrons. 

From these results, it also emerges that the interference between F1 and F2
is destructive since the total cross section is smaller than the
individual cross sections. This makes the backward term dominate,
resulting in the final observation that neutrons are more energetic
when all four contributions are summed, taking into account the
interferences.

As this detailed analysis shows, the comparison with other models, such as
the Valencia model, should consider the distinct physics assumptions
and modeling choices inherent to each approach. The differences
observed underscore the complexity of the underlying nuclear dynamics
and the importance of refining theoretical models to capture the
intricacies of neutrino-nucleus interactions.

Finally, it is important to emphasize that this last discussion does not
contradict the analysis performed to derive Eq. (107), which was based
on the dominance of the $\Delta$ forward process in pp emission. The case
of pp emission differs significantly from pn emission, and the various
contributions enter into a different combination due to isospin
considerations. On the other hand, these arguments cannot be directly
applied to the case of antineutrinos in Fig. 13, where the pn
distribution appears to be fairly symmetric. This symmetry in the
antineutrino case arises from the subtraction of transverse terms in
the hadronic tensor due to the negative sign in Eq. (3) when
contracting the leptonic tensor. Therefore, careful consideration is
needed when assessing the importance of different terms, as it is less
straightforward and may depend on the specific kinematics involved.

\section{Conclusions}

We have explored the semi-inclusive two-nucleon emission reaction
induced by neutrinos and antineutrinos within the framework of the
relativistic mean field of nuclear matter. Our approach involves a
factorization approximation, where the reaction is described by an
elementary two-nucleon cross section multiplied by an integrated
two-hole spectral function. The 2p2h excitations are modeled using a
relativistic treatment of meson-exchange currents.

One notable contribution of this work is the derivation of a simple
analytical formula for the integrated two-hole spectral function. This
formula not only streamlines the calculation of the cross section in
the factorized case but also facilitates a clear interpretation of the
obtained results. To validate the factorized approximation, we have
performed comparisons with exact results obtained through numerical
integration over the angles of one hole in the center-of-mass system
of the two holes.

Our study has also provided a reliable prescription for the elementary
two-nucleon hadronic tensor. This was achieved by evaluating the
tensor at averaged hole momenta that satisfy energy-momentum
conservation. These averaged momenta are chosen to be perpendicular to
the missing momentum in the center-of-mass system of the two holes, as
well as perpendicular to the momentum transfer. This prescription
ensures a consistent treatment of the elementary two-nucleon process
within the factorized model.

Our results demonstrate the efficacy of the factorized model in
capturing essential features of the semi-inclusive cross section,
particularly when considering the angular dependence of the
two-nucleon emission.  The semi-inclusive two-nucleon emission
results, integrated over the energy of one of the particles, exhibit a
remarkable agreement with shell model calculations of
Ref. \cite{Cuy17}. This agreement is noteworthy, especially
considering the Fermi gas nature of our approach. We attribute this
success to the integration over holes in our model, contrasting with
shell models that sum over occupied states, leading to a similar
smearing effect. The correct energy balance, incorporating the
effective mass and vector energy within the relativistic mean field
(RMF), further contributes to the agreement. Our results also show the
dominance of final state configurations close to back-to-back
nucleons, i.e., the angle between is larger to 90 degrees.  A
comparison with a pure phase-space model reveals clear differences in
the final particle distributions, underscoring the importance of
considering the dependence of the hadronic tensor on the 2p2h momenta
in such reactions.

Additionally, we have computed the cross section for neutrinos and
antineutrinos, integrated over the muon kinematics and nucleon angles,
as functions of the outgoing momenta \(p'_1\) and \(p'_2\). The
factorized model significantly simplifies the computational effort,
yielding smooth and distinct distributions.  Our analysis of the
emission distributions for pp and pn pairs has been interpreted in
light of the dominance of the $\Delta$ current. Comparisons with the
Valencia model reveal clear disparities, highlighting the impact of
different model ingredients on the results. These differences
underscore the importance of a detailed understanding of the
underlying physics in neutrino-induced reactions.

This work lays the foundation for future developments that can enhance
our understanding of two-nucleon emission reactions induced by
neutrinos. One avenue for improvement is the incorporation of
short-range correlations, considering that the two initial nucleons
are correlated. A possible approach is to solve the Bethe-Goldstone
equation for the initial state of the two nucleons \cite{Cas23b}, revealing
high-momentum components that facilitate the emission of two
nucleons \cite{Mar23a}. This, in conjunction with the meson-exchange current (MEC)
model, would introduce an interference between short-range
correlations and MEC, adding further complexity and richness to the
reaction dynamics.

In this study, we have neglected the interaction in the final
state. Future work could explore the inclusion of final-state
interactions, providing a more comprehensive description of the entire
reaction process. Another avenue for future research is to incorporate
realistic two-hole spectral functions, akin to those found in the
literature \cite{Geu96,Ben99}.
This would refine the model with a more realistic distribution,
and allow for a more detailed
comparison with experimental data.

In summary, the factorized model developed in this work serves as a
versatile tool for investigating semi-inclusive two-nucleon emission
reactions in neutrino and antineutrino interactions. The insights
gained from this study open up avenues for extending the model to
include more sophisticated physics, such as short-range correlations
and realistic spectral functions, to provide a more accurate
representation of the underlying nuclear dynamics. These advancements
will contribute to the ongoing efforts to unravel the intricacies of
neutrino-induced reactions on nuclear targets.

\appendix

\section{Calculation of the semi-inclusive 2p2h
hadronic tensor in the center of mass system of the two holes}

\label{appa}

In this appendix we reduce an integral of the kind
 \begin{equation} \label{integral}
I(\nH,E)
\equiv 
\int
\frac{d^3h_1}{2E_1}
\frac{d^3h_2}{2E_2} 
 f(\nh_1,\nh_2)
\delta(E_1+E_2-E)
\delta(\nh_1+\nh_2-\nH) 
\end{equation}
to an integral over the relative angles of $\nh_1$ in the CM system of
the two holes with momenta $\nh_1,\nh_2$ and mass $m$. 
Here $ f(\nh_1,\nh_2)$ is an arbitrary function.

We proceed in several steps:

\begin{enumerate}
\item First we prove the inequality
\begin{equation} \label{mmenor}
m^2\leq E_1E_2-\nh_1\cdot\nh_2.
\end{equation}
In fact,
\begin{eqnarray} \label{ineq1}
0 \leq (\nh_1-\nh_2)^2 = h_1^2+h_2^2-2\nh_1\cdot\nh_2
\Longrightarrow 2\nh_1\cdot\nh_2 \leq h_1^2+h_2^2.
\end{eqnarray}
On the other hand we have
\begin{equation} \label{ineq2}
(\nh_1\cdot\nh_2)^2\leq h_1^2h_2^2.
\end{equation}
Combining (\ref{ineq1}) and (\ref{ineq2}),
\begin{eqnarray}
 (\nh_1\cdot\nh_2)^2 +2 (\nh_1\cdot\nh_2)m^2 
&\leq&  
 h_1^2h_2^2+(h_1^2+h_2^2)m^2 \Longrightarrow
\nonumber \\
 (\nh_1\cdot\nh_2)^2 +2 (\nh_1\cdot\nh_2)m^2 +m^4
& \leq & 
 h_1^2h_2^2+(h_1^2+h_2^2)m^2+m^4 \Longrightarrow
\nonumber \\
 (\nh_1\cdot\nh_2+m^2)^2
&\leq&  
( h_1^2+m^2)(h_2^2+m^2) = E_1^2E_2^2 \Longrightarrow
\nonumber \\
 \nh_1\cdot\nh_2+m^2
&\leq&
E_1E_2.
\end{eqnarray}
This concludes the proof of (\ref{mmenor}).

\item 
If $E^2-H^2< 4m^2$, then $I(\nH,E)=0$.

In fact we note that the product of delta functions inside the integral 
(\ref{integral})
is zero unless
\begin{equation}
\nh_1+\nh_2=\nH,  \kern 2cm E_1+E_2=E.
\end{equation}
This implies that 
\begin{equation} \label{inequality}
E^2-H^2 = 2m^2 +2E_1E_2 - 2\nh_1\cdot\nh_2 > 4m^2.
\end{equation}
The last inequality follows from Eq. (\ref{mmenor}).  Conversely, if
this inequality is not satisfied, then the integral (\ref{integral})
is zero.

\item From step \#2 above, the integral (\ref{integral}) 
can be equivalently expressed as 
 \begin{equation} \label{integral2}
I(\nH,E)
=
\int
\frac{d^3h_1}{2E_1}
\frac{d^3h_2}{2E_2} 
 f(\nh_1,\nh_2)
\delta(E_1+E_2-E)
\delta(\nh_1 +\nh_2-\nH) 
\theta(E^2-H^2-4m^2)
\end{equation}

\item
The integral $I(\nH,E)$ can be written in the equivalent form
\begin{eqnarray}
I(\nH,E)&=&
\int d^4h_1 d^4h_2 \delta^4(h_1^\mu+h_2^\mu-H^\mu)f(\nh_1,\nh_2)
\nonumber\\
&& \mbox{}\times
\delta(h_{1\mu}h_1^\mu-m^2)\theta(h_1^0)
\delta(h_{2\mu}h_2^\mu-m^2)\theta(h_2^0)
\theta(E^2-H^2-4m^2) 
\label{invariant}
\end{eqnarray}
where we have introduced the four-vectors $h_1^\mu=(h_1^0,\nh_1)$,
$h_2^\mu=(h_2^0,\nh_2)$, and
 $H^\mu=(E,\nH)$.

To prove the formula, we just need to use the following result from
special relativity:
\begin{equation} \label{invariant2}
\int
\frac{d^3h_1}{2E_1}g(E_1,\nh_1)
=
\int d^4h_1 g(h_1^\mu) \delta(h_{1\mu}h_1^\mu-m^2)\theta(h_1^0),
\end{equation}
where $g(E_1,\nh_1)$ is an arbitrary function.

\item We perform the integral in the CM system of the two-holes 
that moves with momentum $\nH$. We 
 change  variables:
\begin{equation}
 h_1^\mu=\Lambda^\mu{}_\nu h''_1{}^\nu, \kern 1cm
 h_2^\mu=\Lambda^\mu{}_\nu h''_2{}^\nu, \kern 1cm
\end{equation}
where $\Lambda$ is a boost transformation matrix.  Double primed
variables refer to the CM system. It is defined so that the coordinates of
the four vector $H^\mu$ in the moving system are
\begin{equation}
\nH''=0 ,\kern 2cm E''=\sqrt{E^2-H^2}.
\end{equation}
Thus the CM system moves with velocity $\nv= \nH/E$.  In fact the new
component of $H^\mu$ in the direction of $\nv$ is given by the
two-dimensional Lorentz transformation
\begin{equation}
H''= \gamma( H-vE), \kern 2cm \gamma=1/\sqrt{1-v^2},
\end{equation}
but $H''=0$ implies $v=H/E$.  Note that the result of
Eq. (\ref{inequality}) implies $v<1$, so the boost is always
possible.

\item Since $\det \Lambda^\mu{}_\nu=1$ we have
\begin{equation}
\delta^4(h_1^\mu+h_2^\mu-H^\mu)=
\delta^4(\Lambda^\mu{}_\nu(h''_1{}^\nu+h''_2{}^\nu-H''{}^\nu))
=\delta^4(h''_1{}^\mu+h''_2{}^\mu-H''{}^\mu).
\end{equation}
Then we can write the integral (\ref{invariant}) in the CM system 
and using again Eq. (\ref{invariant2}) we arrive to the result
 \begin{equation} \label{integral2}
I(\nH,E)
\equiv 
\int
\frac{d^3h''_1}{2E''_1}
\frac{d^3h''_2}{2E''_2} 
 f(\nh_1,\nh_2)
\delta(E''_1+E''_2-E'')
\delta(\nh''_1+\nh''_2) 
\theta(E^2-H^2-4m^2)
\end{equation}
Integrating over $\nh_2''$ we have $\nh_2''=-\nh_1''$ and
$E''_2=E''_1$. Therefore
 \begin{equation} \label{integral3}
I(\nH,E)
\equiv 
\int
\frac{d^3h''_1}{4(E''_1)^2}
 f(\nh_1,\nh_2)
\delta(2E''_1-E'')
\theta(E^2-H^2-4m^2)
\end{equation}

\item To finish we integrate over the energy $E''_1$ using  
\begin{eqnarray}
E_1''dE_1''=h_1''dh_1'' &\Longrightarrow& d^3h_1''=
h_1''E_1''dE_1''d\Omega_1'',
\\
 \delta(2E''_1-E'') &=& \frac12 \delta(E''_1-\frac{E''}2),
\end{eqnarray}
we obtain the result $E''_1=E''/2$ and 
 \begin{equation} \label{integralfin}
I(\nH,E)
=
\frac{1}{4}
\theta(E^2-H^2-4m^2)
\frac{h_1''}{2E_1''}
\int
d\Omega''_1
 f(\nh_1,\nh_2)
\end{equation}
where $d\Omega_1''=d\cos\theta_1''d\phi_1''$ and
$(\theta''_1,\phi''_1)$ are the angles of $\nh''_1$ in spherical
coordinates.

\end{enumerate}

\section{Integration Limits of  $G(E,H)$}
\label{appb}

In this appendix, we obtain the integration limits of the function
$G(E,H)$ given by Eq. (\ref{g4}), as an integral over the energy $E_1$
of the first hole. 
\begin{equation}
G(E,H) = \frac{2 \pi (m_N^*)^2}{H}
\int_{m_N^*}^{E_F} dE_1   
\theta(E-E_1 - E_{H-h_1})  
\theta(E_{H+h_1}-E+E_1)
\theta(E_F-E+E_1).
\label{g4bis}
\end{equation}
$E_1$ is subjected to the following conditions imposed by
the step functions inside the integral
\begin{eqnarray}
E_{H-h_1}< E-E_1 &<& E_{H+h_1} \label{limite1}, \\
 E-E_1 &<& E_F.  \label{limite2}
\end{eqnarray}
By squaring the first  inequality (\ref{limite1})
and rearranging terms,
\begin{eqnarray}
(m_N^*)^2 +(H-h_1)^2 < (E-E_1)^2 < (m_N^*)^2 + (H+h_1)^2 \Longrightarrow  
\nonumber \\
H^2  - 2 Hh_1  < E^2 - 2EE_1<   H^2 +2Hh_1 \Longrightarrow \nonumber \\
 - 2 Hh_1  < E^2 - 2EE_1 - H^2<   2Hh_1. 
 \end{eqnarray}
Therefore
 \begin{equation}  \label{ineq1}
  |E^2-H^2 - 2 E E_1| < 2 H h_1. 
\end{equation}
Note that $h_1$ also depends on the integration variable $E_1$ so we
need to manipulate the inequality $\ref{ineq1}$ to obtain a condition
involving only $E_1$. It is convenient to 
rewrite the previous equation in terms of
the dimensionless variables normalized with the nucleon mass as
defined in equations (\ref{adimen1},\ref{adimen2}), we have:
\begin{equation}
|\tau + \lambda\epsilon| < \kappa \eta \nonumber \\
\end{equation}
The next step is to take the square of this inequality, 
and using $\eta^2=\epsilon^2-1$ and $\kappa^2-\lambda^2=\tau$
\begin{eqnarray}
\tau^2+\lambda^2\epsilon^2+2\tau\lambda\epsilon < \kappa^2\eta^2
=\kappa^2(\epsilon^2-1) \Rightarrow
\nonumber\\
\tau^2+2\tau\lambda\epsilon < (\kappa^2-\lambda^2)\epsilon^2-\kappa^2=
\tau\epsilon^2-\kappa^2.
\end{eqnarray}
Moving the terms that depend on $\epsilon$ to the right-hand side of
the inequality.
\begin{eqnarray}
\tau^2+\kappa^2 
&<&  \tau(\epsilon^2 -2\lambda \epsilon)
\nonumber\\
&=&\tau[ (\epsilon-\lambda)^2 -\lambda^2]
\nonumber\\
&=&\tau[ (\epsilon-\lambda)^2 -\kappa^2+\tau]
\nonumber\\
&=&\tau(\epsilon-\lambda)^2 -\tau\kappa^2+\tau^2.
\end{eqnarray}
Therefore we can write
\begin{equation}
\kappa^2(1+\tau) <  \tau(\epsilon-\lambda)^2 
\end{equation}
Finally, we divide by the variable $\tau$, taking into account that $\tau < 0$,
\begin{eqnarray}
(\epsilon-\lambda)^2 &<& \kappa^2\left(1+\frac{1}{\tau}\right) \Longrightarrow
\nonumber\\
|\epsilon-\lambda| &<& \kappa\sqrt{1+\frac{1}{\tau}}.
\end{eqnarray}
This implies that $\epsilon$ is in the interval
\begin{equation}
\lambda-\kappa\sqrt{1+\frac{1}{\tau}}
< \epsilon < 
\lambda+\kappa\sqrt{1+\frac{1}{\tau}}.
\label{condicion1}
\end{equation}

Now let's examine the restrictions imposed by the conditions that the
energies of the holes, $E_1$ and $E_2=E-E_1$, must be greater than the mass and
less than the Fermi energy.
\begin{eqnarray*}
m_N^* < E-E_1 < E_F, \\
m_N^* < E_1 < E_F,
\end{eqnarray*}
or in terms of dimensionless variables
\begin{eqnarray}
1< 2\lambda-\epsilon < \epsilon_F \Longrightarrow
 2\lambda-\epsilon_F < \epsilon 
\label{condicion2}\\
1< \epsilon < \epsilon_F. \label{condicion3}
\end{eqnarray}
For all three conditions (\ref{condicion1}--\ref{condicion3})
to be fulfilled simultaneously, $\epsilon$ must lie in the following interval.
\begin{equation}
\epsilon_A < \epsilon < \epsilon_B,
\end{equation}
where the lower and upper limits are given by
\begin{eqnarray}
\epsilon_A 
&=& 
\mbox{Max} 
\left\{ 
\lambda-\kappa\sqrt{1+\frac{1}{\tau}},\,
2\lambda-\epsilon_F,\, 1
\right\}
\\
\epsilon_B
&=& 
\mbox{min} 
\left\{ 
\lambda+\kappa\sqrt{1+\frac{1}{\tau}},\,
\epsilon_F
\right\}.
\end{eqnarray}

\section*{Acknowledgments}
The Work was supported by Grant No. PID2020-114767GB-I00 funded by
MCIN/AEI /10.13039 /501100011033; and by Grant No. FQM-225 funded by
Junta de Andalucia

\end{document}